\documentclass[twocolumn,superscriptaddress,amsmath,amstex,amssymb,citeautoscript]{revtex4-2}

\usepackage{graphicx}
\usepackage{physics}
\usepackage{etoolbox}
\usepackage{hyperref}
\usepackage{times}
\usepackage{bm}
\usepackage{makecell}
\usepackage{color}
\usepackage{amssymb}
\usepackage{amsmath}
\usepackage{braket}
\usepackage{outline}
\usepackage[table]{xcolor}

\begin{document}

\title{Scalable Spin Squeezing from Finite Temperature Easy-plane Magnetism}

\author{
\normalsize{Maxwell Block$^{*,1}$, Bingtian Ye$^{*,1}$, Brenden Roberts$^1$, Sabrina Chern$^1$, Weijie Wu$^4$, Zilin Wang$^4$,} \\
\normalsize{Lode Pollet$^{2,3}$, Emily J. Davis$^4$, Bertrand I. Halperin$^1$, Norman Y. Yao$^{1,4,\dag}$} \\
\footnotesize{$^1$Department of Physics, Harvard University, Cambridge, MA 02139, USA}\\
\footnotesize{$^2$Arnold Sommerfeld Center for Theoretical Physics, University of Munich, Theresienstr. 37, 80333 München, Germany}\\
\footnotesize{$^3$Munich Center for Quantum Science and Technology (MCQST), Schellingstr. 4, 80799 München, Germany}\\
\footnotesize{$^4$Department of Physics, University of California, Berkeley, CA 94720, USA}\\
\footnotesize{$^*$These authors contributed equally to this work.}
\footnotesize{$^\dag$Corresponding author nyao@fas.harvard.edu}
}






\begin{abstract}
Spin squeezing is a form of entanglement that reshapes the quantum projection noise
to improve measurement precision.
Here, we provide numerical and analytic evidence for the following conjecture: any Hamiltonian exhibiting finite temperature, easy-plane ferromagnetism can be used to generate scalable spin squeezing, thereby enabling quantum-enhanced sensing. 
Our conjecture is guided by a connection between the quantum Fisher information of pure states and the spontaneous breaking of a continuous symmetry. 
We demonstrate that spin-squeezing exhibits a phase diagram with a sharp transition between scalable squeezing and non-squeezing. 
This transition coincides with the equilibrium phase boundary for XY order at a finite temperature. 
In the scalable squeezing phase, we predict a sensitivity scaling that lies in between the standard quantum limit and the scaling achieved in all-to-all coupled one-axis twisting models. 
A corollary of our conjecture is that short-ranged versions of two-axis twisting cannot yield scalable metrological gain. 
Our results provide insights into the landscape of Hamiltonians that can be used to generate metrologically useful quantum states.
\end{abstract}

\maketitle

\emph{Introduction.}---Quantum enhanced metrology makes use of many-body entangled states to perform measurements with greater precision than would be possible using only classically correlated particles~\cite{giovannetti_advances_2011,degen_quantum_2017,schnabel_quantum_2010,wasilewski_quantum_2010,leibfried_toward_2004}.
Identifying states suitable for quantum metrology is a delicate challenge: nearly all states in 
Hilbert space are highly entangled, but almost none of them exhibit the structured entanglement required for enhanced sensing.
Indeed, only a handful of metrologically useful quantum states are typically discussed, e.g.~GHZ states~\cite{bouwmeester_observation_1999, pan_experimental_2000, omran_generation_2019, pogorelov_compact_2021, jones_magnetic_2009}, Dicke states~\cite{dicke_coherence_1954, pezze_quantum_2018, wieczorek_experimental_2009, lucke_detecting_2014, zou_beating_2018}, and squeezed states~\cite{kitagawa_ueda, wineland_squeezed_1994, hosten_measurement_2016, esteve_squeezing_2008, appel_mesoscopic_2009, muessel_scalable_2014}. 
Identifying universal principles for adding to this list remains an important challenge, especially in the context of efficiently preparing metrologically useful states from un-entangled product states. 

 \begin{figure*}
	\centering\includegraphics[width=0.8\textwidth]{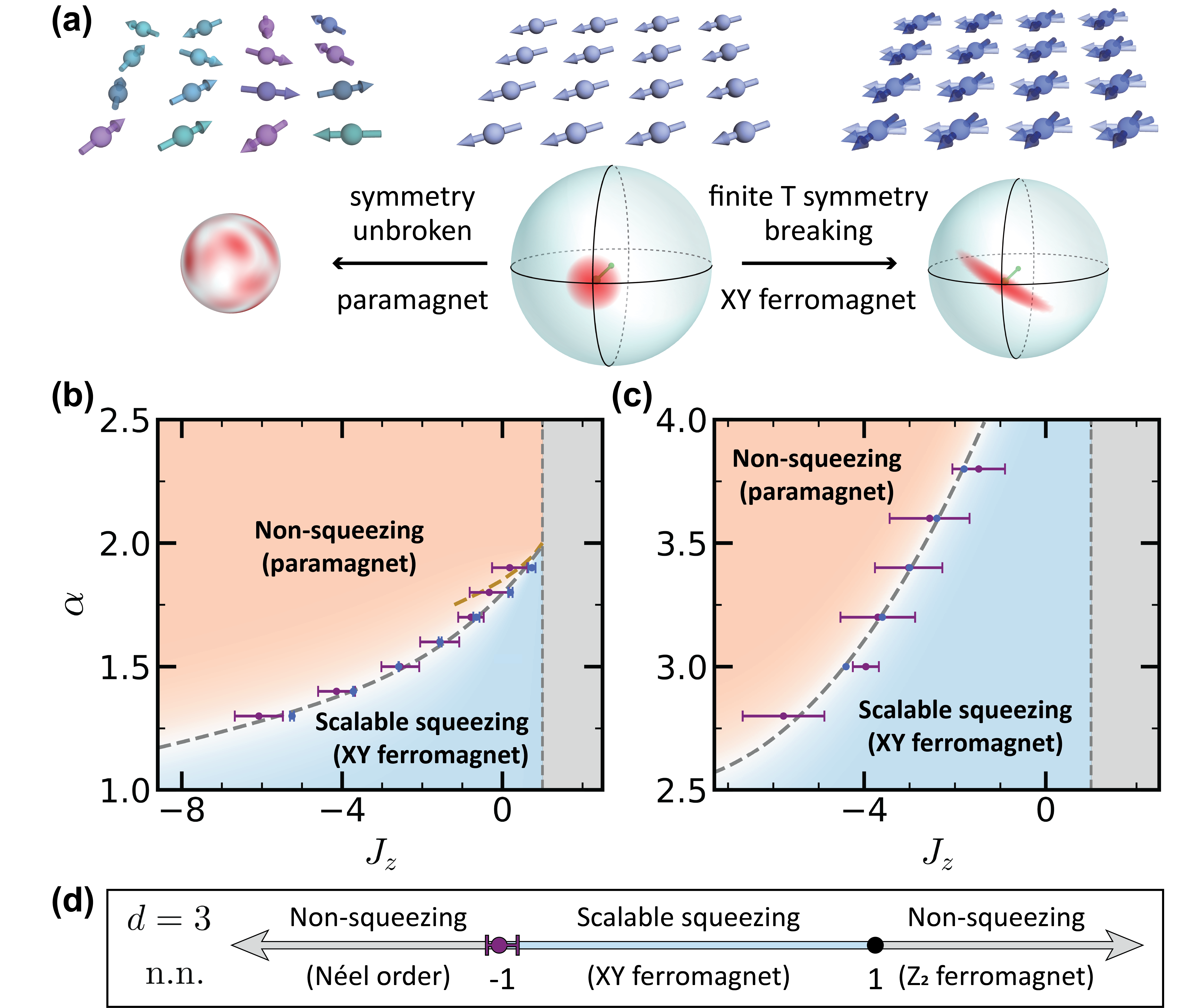}
\caption[Scalable spin squeezing from easy-plane ferromagnetism]{(a) Illustration of the connection between spin squeezing and XY ferromagnetic order.
The central panel shows an initial product state polarized in the equatorial plane, with a schematic of its quantum projection noise (i.e.~spin Husimi-Q distribution) indicated in red. 
When the effective temperature of the initial state is below the critical temperature for $U(1)$ symmetry breaking (right panel), the system equilibrates to an XY ferromagnet and exhibits scalable squeezing. 
When the initial state is above the critical temperature, the system equilibrates to a paramagnet, where the total spin length vanishes, precluding
spin squeezing (left panel).
(b,c) Phase diagrams for scalable spin squeezing as a function of the XXZ anisotropy, $J_z$, and the interaction power-law, $\alpha$, in dimensions $d=1,2$.
The location of the squeezing phase transition is computed via DTWA (purple markers). 
The location of the ordering transition (blue markers, dashed line guide to the eye) is computed in 1D via imaginary time evolution on matrix product operators and in 2D via finite temperature quantum Monte Carlo.
In both cases, the squeezing and order phase boundaries are in close agreement.
The gold dashed line in (b) is an analytic approximation of the phase boundary from spin-wave theory \cite{supp}.
(d) Phase diagram for scalable squeezing as a function of $J_z$ for a 3D nearest-neighbor interacting model. For $\abs{J_z} > 1$, there is N\'eel or Ising order (rather than CSB), so squeezing does not occur.
In (b,c,d), the squeezing critical point (errorbar) is given as the center (width) of the interval containing all squeezing critical points from 25 resamples of DTWA results.
	}
	\label{fig:1}
\end{figure*}
 
One such principle, which is particularly powerful, stems from the observation that the quantum Fisher information (QFI) for pure states is fundamentally connected to spontaneous symmetry breaking. 
On the one hand, the QFI characterizes the maximum sensitivity of a given quantum state, $|\psi\rangle$, to a specific perturbation, $\hat{\mathcal{O}}=\sum_i \hat{\mathcal{O}}_i$  and  simply reduces to the variance of $\hat{\mathcal{O}}$ \cite{braunstein_statistical_1994}.
On the other hand, spontaneous symmetry breaking (SSB) leads to the existence of long-range connected correlations of the order parameter $\mathcal{O}$ (provided that $\mathcal{O}$ does not commute with the Hamiltonian) and thus a variance which scales quadratically in system size (see Methods for a detailed discussion). 
%
Thus, any pure quantum state exhibiting long-range connected correlations can be utilized to perform Heisenberg-limited sensing.
%
Combining these considerations with the eigenstate thermalization hypothesis---which asserts that generic quantum dynamics cause few-body observables to reach thermal equilibrium \cite{deutsch_quantum_1991, srednicki_chaos_1994, rigol_thermalization_2008}---suggests a broad and simple guiding principle for preparing metrologically useful states from product states: Identify a Hamiltonian, $H$, exhibiting finite temperature order. 
Then, find an unentangled state $\ket{\psi_0}$ whose effective temperature is below $T_\textrm{c}$ and time evolve.
The effective temperature, $T$, of the initial state $|\psi_0 \rangle$ is defined by its energy density, such that 
$\Tr[\rho\;H] = \bra{\psi_0} H \ket{\psi_0}$, where $\rho = \frac{e^{-H/T}}{\mathcal{Z}}$ is a thermal density matrix. 

The above strategy for generating metrologically useful states may seem surprising. 
Indeed, seminal work exploring the relationship
between the QFI and equilibrium phases of matter has come to the opposite conclusion: Thermal density matrices cannot be used to perform metrology beyond the standard quantum limit~\cite{hauke_measuring_2016,gabbrielli_multipartite_2018}.
This limitation applies even for thermal states below the critical temperature, $T_c$, for spontaneous symmetry breaking.
The key ingredient which allows one to escape this thermal bind is that the non-equilibrium quench can generate coherences which are not present in the thermal state. 
Thus, despite having the same energy density, the quenched state and the thermal density matrix can exhibit extremely different metrological properties  (see Methods).

The efficiency of our proposed strategy depends crucially on the nature of the symmetry being broken. For discrete symmetries, it takes an exponentially long time (in system size) to develop a large QFI (see Methods).
For continuous symmetries however, large QFI can develop significantly faster (i.e.~in polynomial time)~\cite{chen_continuous_2022, roscilde_cat_states}.
This principle suggests that \emph{any} system with finite-temperature continuous symmetry breaking can be used to prepare states for quantum enhanced sensing.
However, validating (or refuting) this general statement is challenging.
To this end, we apply the above principle to the simplest and most experimentally relevant case of $U(1)$ symmetry breaking, and provide extensive numerical and analytic evidence for the following remarkable conjecture: Finite-temperature easy-plane ferromagnetism (i.e. XY magnets) enables the preparation of states with large QFI, specifically in the form of scalable spin squeezing.

Our main results are threefold. 
First, we establish a phase diagram for spin squeezing (Fig.~\ref{fig:1}), with a sharp transition distinguishing scalable squeezing from non-squeezing. 
Second, we argue that this transition occurs precisely when the effective temperature of the initial state $\ket{\psi_0}$ equals the critical temperature for continuous symmetry breaking (CSB). 
Finally, we show that the squeezing manifests a novel scaling with system size---whose origin is extremely subtle---that leads to a phase sensitivity $\sim N^{^{-\frac{7}{10}}}$, between the standard quantum limit ($\sim N^{^{-\frac{1}{2}}}$) and the Heisenberg limit ($\sim N^{-1}$).
Intriguingly, for parametrically long time-scales in the inverse temperature of the initial state, we observe a sensitivity scaling as $\sim N^{^{-\frac{5}{6}}}$, matching that achieved in all-to-all coupled easy-plane spin models, i.e.,~so-called one-axis twisting (OAT) models ~\cite{kitagawa_ueda}. 
A corollary of our conjecture is that short-ranged (i.e.~not all-to-all connected) generalizations of the two-axis twisting model cannot yield a scalable metrological gain, a fact which we numerically verify (see Methods).
We note that several prior works have explored aspects of the connection between scalable spin-squeezing and $U(1)$ symmetry breaking \cite{perlin,comparin,comparin_scalable_2022}.
This connection has been viewed from the perspective of zero temperature, ground-state symmetry breaking \cite{comparin,comparin_scalable_2022}. 
While this perspective yields invaluable intuition, it cannot fully explain how squeezing arises in a general (finite-temperature) quench.

In contrast, our results are based on the principle that after a short initial period of time, the long-wavelength, low-frequency properties of the system can be described by hydrodynamic equations involving the conserved $z$-component of the spin-density and the orientation of the magnetization in the $x$-$y$ plane \cite{halperin_hydrodynamic_1969}. 
While the form of these hydrodynamical equations is fixed, their  parameters depend on the microscopic Hamiltonian; therefore, we utilize a variety of approximate numerical methods to investigate both the squeezing dynamics as well as the equilibrium phase diagram of a broad class of easy-plane spin models.

 %
 %

\emph{Connecting squeezing to XY magnetism}.---To investigate our conjecture, let us consider the paradigmatic family of $U(1)$-symmetric spin Hamiltonians: the long-range $S=1/2$ XXZ model
\begin{equation}
H_{\textrm{XXZ}} = -\sum_{i<j} \frac{J_\perp(\sigma^x_i \sigma^x_j + \sigma^y_i \sigma^y_j) + J_z \sigma^z_i \sigma^z_j}{r_{ij}^\alpha},
\end{equation}
where $J_\perp/J_z$ characterizes the easy-plane anisotropy and $\alpha > d$ is the long-range power law. 
This class of Hamiltonians is the most natural and widely studied generalization of one-axis twisting \cite{perlin, comparin, cappellaro_quantum_2009, davis_protecting_2020, kwasigroch_bose-einstein_2014, kwasigroch_synchronization_2017, rey_many-body_2008} and is also realized in a number of quantum simulation platforms ranging from solid-state spins and optical-lattice Hubbard models to  ultracold polar molecules and Rydberg atom arrays~\cite{davis_probing_2021, greiner_quantum_2002, gross_quantum_2017, ni2008polarmolecule, chen_continuous_2022, chomaz_long-lived_2019, tran_quantum_2016}. 
While our conjecture applies in all dimensions,we note that in $d=1,2$ [Fig.~\ref{fig:1}(b,c)] finite-temperature continuous symmetry breaking is only possible for sufficiently long-range power laws whereas in $d=3$ [Fig.~\ref{fig:1}(d)] it is possible even with nearest-neighbor interaction \cite{fn3, bruno2001absence, maghrebi_continuous_2017}.
To establish the connection between squeezing and order we utilize a variety of numerical tools: the discrete truncated Wigner approximation (DTWA) for spin-squeezing dynamics,
imaginary time evolution of matrix product operators for diagnosing finite-temperature $U(1)$ symmetry breaking order in $d=1$, and finite-temperature path integral quantum Monte Carlo for diagnosing order in $d=2,3$ \cite{schachenmayer, hauschild_efficient_2018}.
In addition, whenever possible, our numerical results are carefully benchmarked with a combination of time-dependent variational Monte Carlo, Krylov subspace methods, and exact diagonalization \cite{supp}.


\begin{figure*}
	\hspace{-10mm}
\includegraphics[width=0.95\textwidth]{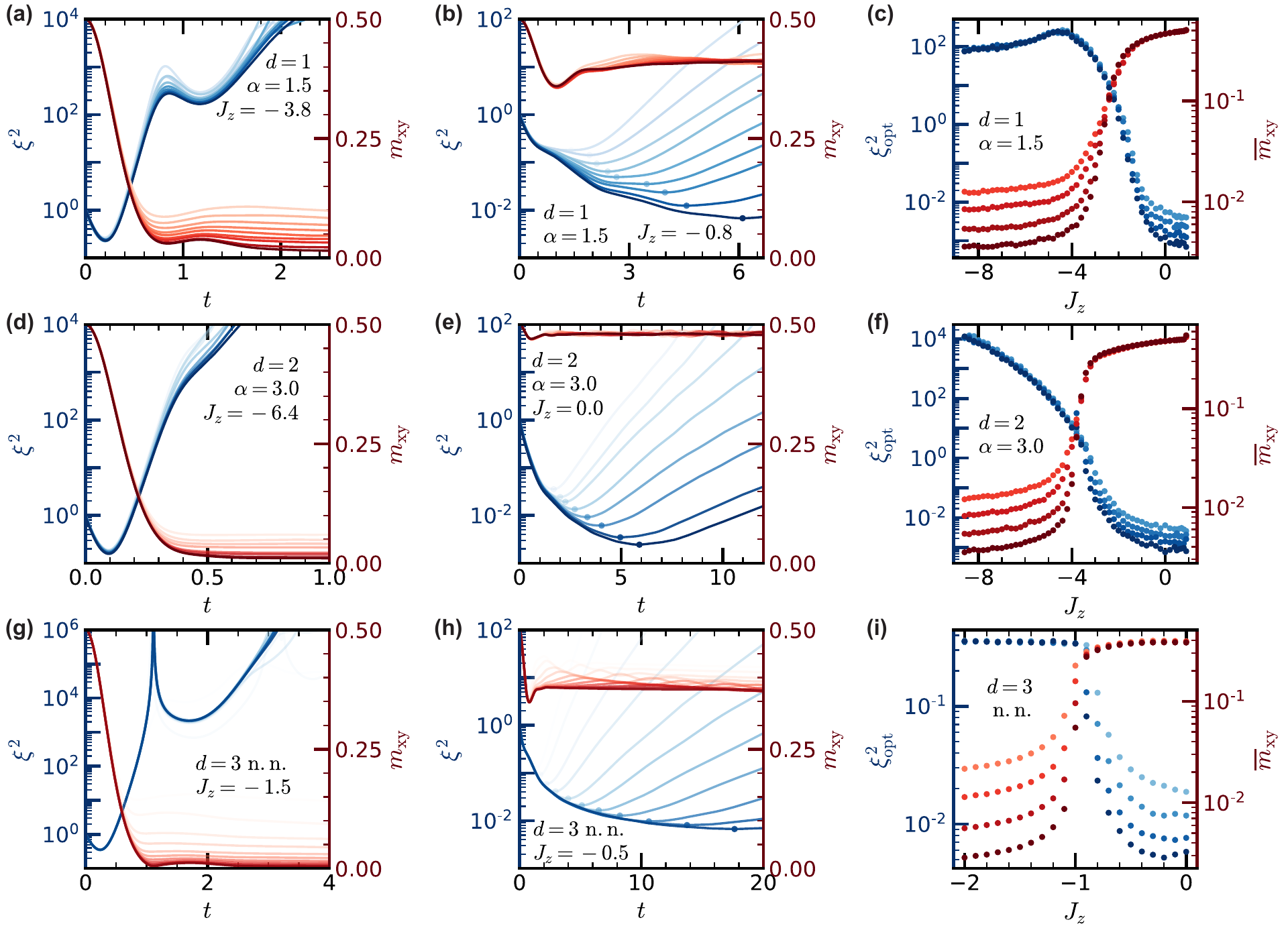}
	\caption[Dynamics of easy-plane order and squeezing]{(a,b) Depicts the dynamics ($d=1$) of the squeezing parameter $\xi^2$ (blue) and the XY order, $m_{\textrm{xy}}$ (red), as a function of time, upon quenching from an initial product state polarized in the $\hat{x}$ direction. 
Opacity increases with system size (from $N=10^2 - 10^4$).
In the paramagnetic phase (a), $m_{\textrm{xy}}$ decays to zero, while the squeezing parameter does not improve with system size.
In the ferromagnetic phase (b), $m_{\textrm{xy}}$ plateaus to a finite value, and the squeezing parameter scales with system size. 
(d,e) and (g,h) Depict the analogous dynamics for $d=2$ and $d=3$, with system sizes $N=10^2 - 10^4$ and $N=10^2-10^5$. 
(c) Fixing $\alpha =1.5$ in $d=1$, the optimal squeezing parameter (blue) and the plateau value of the XY order (red) are shown as a function of $J_z$.
Darker color shades correspond to larger system sizes (with $N=5 \cdot 10^3 - 5 \cdot 10^4$).
In the paramagnetic phase (large $J_z$ magnitude), the XY order decays with increasing system size, while in the ferromagnetic phase the XY order is independent of system size, indicating a phase transition at $J_c \approx -2.4$.
The behavior of the optimal squeezing exhibits a separatrix at the same value of $J_c$, where it transitions from being system-size independent to scaling with $N$.
(f) and (i) Depict analogous $J_z$-cuts in the phase diagrams for $d=2$ fixing $\alpha = 3.0$ (with $N=5 \cdot 10^3 - 5 \cdot 10^4$) and $d=3$ (with $N=2 \cdot 10^3 - 10^5$).
Calculations were performed using the DTWA approximation.
	}
	\label{fig:2}
\end{figure*}

Let us begin in $d=1$ with ferromagnetic XY interactions (hereafter, we set $J_\perp=1$).
For $J_z > 1$, the Hamiltonian lies in the easy-axis regime and can only exhibit discrete (Ising) symmetry breaking; this immediately rules out the possibility of quantum-enhanced sensing at accessible time scales [Fig.~\ref{fig:1}(b)]. 
For $J_z < 1$, the system can exhibit continuous symmetry breaking at finite temperatures provided that $\alpha < 2$.
Consider the parameter space with weak power laws and strong antiferromagnetic Ising interactions [pink, Fig.~\ref{fig:1}(b)]. 
Taking our initial state as the fully polarized coherent spin state in the $x$ direction, $\ket{x} = | \rightarrow \cdots \rightarrow \rangle$, we evolve under $H_{\textrm{XXZ}}$ and measure both the average XY magnetization, $m_{\textrm{xy}} = [\expval{X^2+Y^2}/N^2]^{1/2}$ and the squeezing parameter $\xi^2 \equiv N \min_{\hat{n}\bot \hat{x}}{\mathrm{Var}[\hat{n}\cdot \vec{S}]}/\langle X\rangle^2 $ (which entails a phase sensitivity of $\Delta \phi = \sqrt{\frac{\xi^2}{N}}$); here, $X = \frac{1}{2}\sum_i \sigma^x_i$ (with $Y$ and $Z$ defined analogously) and $\vec{S} = (X,Y,Z)$.
As a function of increasing system size, the magnetization decays to zero, indicating thermalization to a disordered state [Fig.~\ref{fig:2}(a)]. 
Meanwhile, the squeezing parameter, which quantifies the enhancement in sensitivity over the initial coherent state, exhibits marginal improvement at short times.
However, this improvement \emph{does not} scale with system size and at late times, $\xi^2$ steadily worsens [Fig.~\ref{fig:2}(a)]. 

The dynamics in the opposite parameter space [blue, Fig.~\ref{fig:1}(b)], with strong power laws and weak antiferromagnetic Ising interactions is markedly distinct.
Here, the XY magnetization rapidly equilibrates to a system-size-independent value, providing a signature of robust continuous symmetry breaking [Fig.~\ref{fig:2}(b)].
Accompanying the presence of order is the existence of scalable spin squeezing, where the optimal squeezing (attained after local relaxation, see Methods) improves with system size, (i.e.~$\xi_\textrm{opt}^2 \sim N^{-\nu}$ with $\nu >0$) and occurs at later and later times [Fig.~\ref{fig:2}(b)].
This is precisely reminiscent of the behavior  in the one-axis-twisting model, where $\xi_\textrm{opt}^2 \sim N^{^{-\frac 2 3}}$.


The essence of our conjecture is already captured by the above dichotomy: thermalizing to a disordered state correlates with non-squeezing, while thermalizing to an ordered state correlates with scalable squeezing.
But our conjecture is stronger than claiming an association between squeezing and finite-temperature order; rather, we argue that they are two facets of the same phase.
To more quantitatively investigate this, for each point in parameter space, $\{\alpha, J_z\}$, we extract the optimal squeezing and the late-time XY magnetization as a function of system size (see Methods for details). 
Depicted in Figure \ref{fig:2}(c), is a cut across  parameter space, fixing  $\alpha = 1.5$ and varying $J_z$.
For large, negative $J_z$, the XY magnetization plateau vanishes with increasing system size. As $J_z$ becomes weaker and enters the ferromagnetic regime, there is a clear separatrix --- indicative of a symmetry-breaking phase transition --- where the  value of the magnetization plateau  becomes system-size independent.
The scaling behavior of the optimal squeezing is in perfect correspondence with the magnetization. 
When the  magnetization vanishes, the squeezing does not scale; when it is finite, the optimal squeezing exhibits its own separatrix and pronounced scaling with system size. 

This change in scaling determines the location of the squeezing transition: for each value of $J_z$, we fit $\xi_{\rm opt}^2 \sim N^{-\nu}$ and associate the critical point, $J_\textrm{c}$, with the onset of $\nu \gtrsim 0$ (see Methods).
Repeating this procedure as a function of $\alpha$ leads to the squeezing phase boundary (\ref{fig:1}(b), purple points).
One can also determine the $U(1)$ symmetry-breaking phase boundary, which occurs when the effective temperature of the initial  state, $\ket{x}$, crosses the critical temperature for XY order.
To this end, we cool an infinite temperature matrix product operator (representing the density matrix) to the energy density of the $\ket{x}$-state using imaginary time evolution and perform finite-size scaling analysis to obtain the ordering phase boundary (\ref{fig:1}(b), blue points) ~\cite{supp}.
%
%
The squeezing and ordering phase boundaries coincide within error bars, supporting our conjecture that squeezing emerges from easy-plane magnetism.
%


\begin{figure*}
	\hspace{-7mm}
	\includegraphics[width=0.95\textwidth]{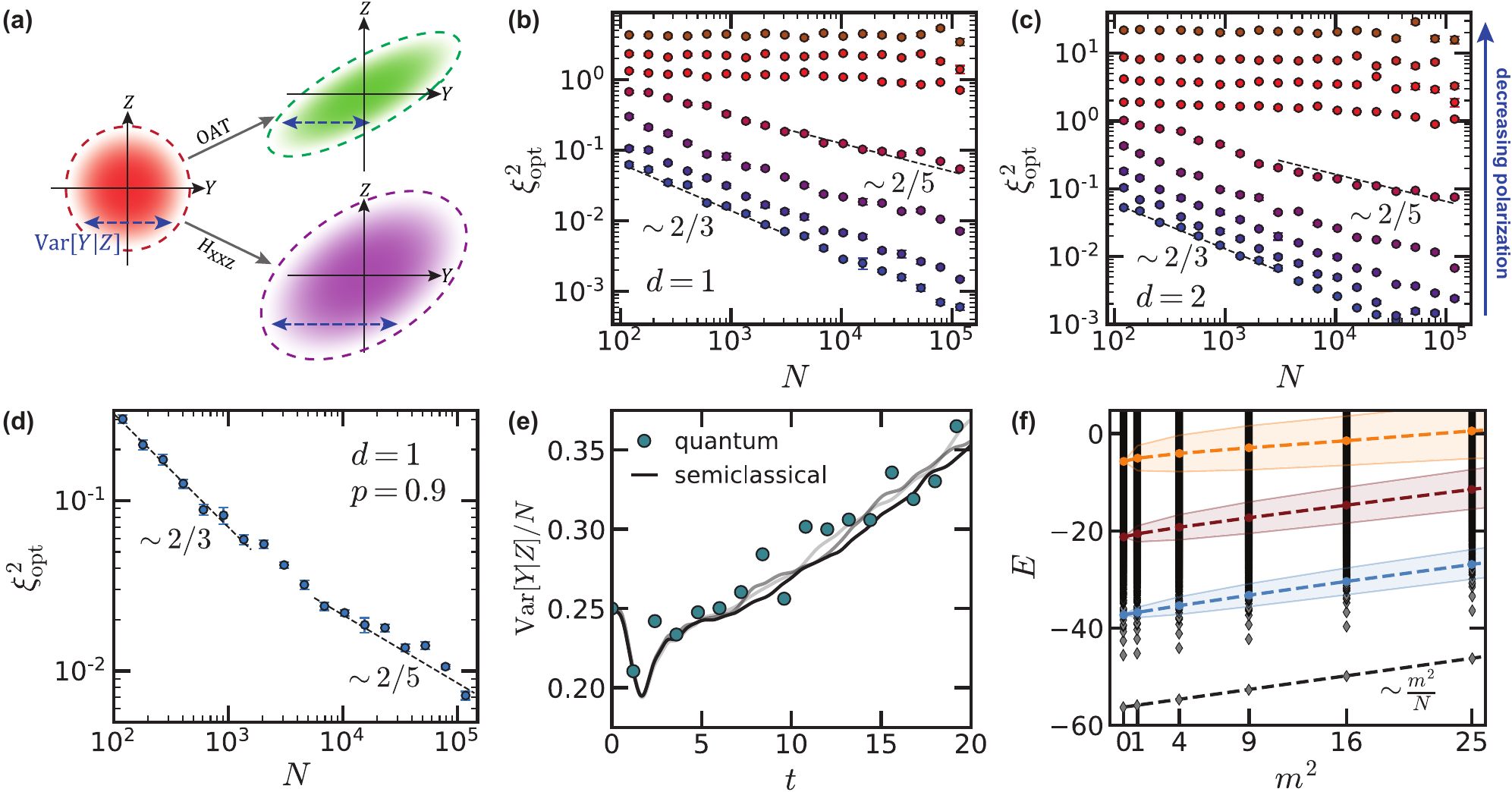}
	\caption[Impact of finite temperature on squeezing]{
	(a) Schematic illustrating the dynamics of conditional variance. 
In the OAT model (green), $\mathrm{Var}[Y|Z]$ remains constant, while in the XXZ model (purple), it exhibits linear growth. 
(b) Optimal squeezing as a function of system size from DTWA simulations. 
The polarization of the initial state decreases from $p=1.0$ (blue) to $p=0.7$ (red); this tunes the effective temperature of the initial state across the critical temperature. 
As the polarization decreases, the scaling  changes sharply from $N^{-2/5}$ to $N^0$. 
At lower effective temperatures ($p\approx 1$), the asymptotic $N^{-2/5}$ scaling emerges only for larger system sizes. 
(c) Analogous results for $d=2$; polarization ranges from $p=1.0$ (blue) to $p=0.6$ (red). 
(d) Squeezing scaling for $d=1$ with polarization $p=0.9$. As the system size increases, $\xi^2_{\rm opt}$ crosses over from $\sim N^{-2/3}$ to the asymptotic prediction $\sim N^{-2/5}$.
(e) In the squeezing phase, both DTWA simulations ($N=1000,2000,4000$ with $Z=0$) and exact quantum dynamics ($N=18$) show linear growth of the conditional variance (also Extended Data Fig.1).
(f) Diamonds depict the spectrum of a $d=1$ $H_\mathrm{XXZ}$ model for $N=24$.
The eigenstates in different $m$-sectors are connected by total-spin raising and lowering operators.
The ground state manifold exhibits a so-called  ``Anderson tower'' structure where the energies scale as $E \sim m^2/N$~\cite{tasaki_long-range_2019}.
At finite temperatures , this manifold becomes a distribution with an approximate
``Anderson tower'' structure: The set of states that are connected to a given initial finite-energy state  upon successive applications of the total-spin raising or lowering operator will have a distribution of energies, which is illustrated here by the blue, red and orange regions for three initial states with $m=0$ but increasing energies $E_0$. The dashed lines, which indicate the mean values of the distributions, exhibit the same scaling, $E-E_0\sim m^2/N$, but the standard error, indicating by shading, increases with $E_0$ .
In all (b-f), $\alpha=1.5$ for $d=1$, $\alpha=3.0$ for $d=2$ and $J_z=0$. In (b-e), $\xi^2_{\rm opt}$ is given as the mean of 25 resamples of DTWA results $\pm$ SD.
	}
	\label{fig:3}
\end{figure*}

Two additional remarks are in order. 
%
%
First, we demonstrate precisely the same correspondence between squeezing and order in $d=2$ [Fig.2(d-f)] and in $d=3$ [Fig.2(g-i)].
%
%
Second, in $d=1,2$, scalable squeezing disappears as $J_z$ decreases because the temperature of the initial $\ket{x}$-state becomes higher than the critical temperature; however, for the nearest-neighbor model in $d=3$, the squeezing disappears for $J_z < -1$, because the continuous-symmetry-breaking XY ferromagnet is replaced by discrete-symmetry-breaking N\'eel order.

Finally, a semiclassical view of the dynamics provides the essential intuition for the connection between squeezing and order.
In this framework, the initial state $\ket{x}$ is viewed as a quasi-probability distribution on the total-spin phase space, represented by a sphere of radius $|S|=N/2$ [middle, Fig.~\ref{fig:1}(a)].
In the all-to-all coupled (OAT) case, squeezing develops as slices from this distribution rotate about the  $\hat{z}$-axis with an angular velocity proportional to total $Z = \frac{1}{2}\sum_i \sigma^z_i$.
Our key insight is that even in the power-law coupled case, or the short-ranged case with $d\geq3$, a similar picture holds as long as the state remains ordered.
Specifically, if the effective temperature of $\ket{x}$ is below the equilibrium CSB critical temperature, then $m_{\textrm{xy}}^2>0$  [Fig~\ref{fig:2}(b,e,h)] and the initial quasi-probability distribution will simply relax, on a microscopic timescale independent of $N$, to a slightly distorted distribution on a smaller phase space of radius $m_{\textrm{xy}} N$ [Fig~\ref{fig:1}(a), right].
This ``dressed" distribution will then evolve qualitatively similarly to the all-to-all coupled case.

 
\emph{Scalable squeezing in the XY magnet}.---Based upon the semi-classical intuition from above, one might naively expect that squeezing in the finite-temperature XY magnet should exhibit the same scaling as in one-axis twisting.
However, this is oversimplified. 
To see this, consider the conditional variance of $Y$ given $Z$, denoted as ${\rm Var}[Y|Z]$, within the semi-classical approximation.
As illustrated in Fig.~\ref{fig:3}(a), for one-axis twisting, ${\rm Var}[Y|Z]$ is a constant of motion: each $Z$-slice of the probability distribution rotates rigidly about the sphere.
But in a system that does not conserve total spin, e.g.~$H_{\textrm{XXZ}}$, the conditional variance will increase as a function of time [Fig.~\ref{fig:3}(a)]; indeed, the hydrodynamic model suggests that $Y|Z$ evolves diffusively, so ${\rm Var}[Y|Z]$ grows linearly in time with a slope that depends strongly on temperature (see Methods).
To understand the impact of this variance growth, we note that (within a semi-classical picture)~:
\begin{equation}
    \xi^2(t) \approx \frac{{\rm Var}[Y|Z]/N}{4 m_\textrm{xy}^4 (\chi t)^2}+\frac{(\chi t)^4}{24m_\textrm{xy}^2 N^2},
\end{equation}
where $\chi$ is the effective one-axis twisting strength, related to the $z$-axis spin susceptibility (see Methods and \cite{kitagawa_ueda, roscilde_cat_states}). When the conditional variance remains constant (i.e.~one-axis twisting),  optimizing over $t$ yields   $\xi^2 \sim N^{^{-\frac{2}{3}}}$ (leading to a phase sensitivity $\sim N^{^{-\frac{5}{6}}}$). However, linear growth of the conditional variance predicts instead that $\xi^2 \sim N^{^{-\frac{2}{5}}}$ (leading to a phase  sensitivity $\sim N^{^{-\frac{7}{10}}}$).

At low temperatures, the slope of the variance growth is small, suggesting that the asymptotic scaling behavior will only be observed at extremely large system sizes.
To control the temperature (without changing the local energy scale of $H$), we introduce an additional tuning parameter: the polarization of the initial state. 
For polarization $p$, we initialize a product state, where each spin points along $+x$ with probability $(1+p)/2$ and along $-x$ with probability $(1-p)/2$.
This polarization tunes the effective temperature of the initial state.
At low temperatures ($p \approx 1$), the squeezing appears to scale as $\sim N^{^{-\frac{2}{3}}}$ [Fig.~\ref{fig:3}(b,c)].
At intermediate temperatures near the transition, the squeezing scales as $\sim N^{^{-\frac{2}{5}}}$ over multiple decades in system size [Fig.~\ref{fig:3}(b,c)].
As soon as the polarization tunes the temperature above $T_\textrm{c}$, a jump occurs in the scaling behavior and $\xi^2$ becomes independent of system size [Fig.~\ref{fig:3}(b,c)].
We note that the low-temperature scaling of $N^{^{-\frac{2}{3}}}$ is not expected to hold as $N\rightarrow \infty$; rather, as shown in Fig.~\ref{fig:3}(d), the scaling will eventually cross over to the asymptotic behavior of $ N^{^{-\frac{2}{5}}}$.

One might naturally wonder whether the above analysis for partially polarized product states could also apply to partially polarized \emph{mixed} initial states?
To be specific, let us consider a density matrix of the form,
 $\rho = \left[ (1+p)/2 \ket{\rightarrow}\bra{\rightarrow} + (1-p)/2\ket{\leftarrow}\bra{\leftarrow}\right ]^{\otimes N}$, which is particularly relevant to experiments~\cite{chen_continuous_2022}.
From a computational perspective, calculating the squeezing parameter for pure states and mixed states is nearly identical, with the only difference being an order of averages.
 In principle, since squeezing is a  nonlinear function of expectation values, this averaging can matter.
 However, we find that the  results for pure and mixed states 
 converge together extremely rapidly with system size, and are essentially identical for $N \gtrsim 10^2$. 
 This leads us to the intriguing conclusion that XY magnetism enables scalable squeezing \emph{even} of certain mixed initial states.
 %
 Most optimistically, one might hope that this applies to \emph{any} mixed initial state with finite polarization---exploring this tantalizing possibility remains a direction for future research.


Let us now return to the distinction between the scaling of spin squeezing in all-to-all versus finite-range interacting systems. Although our semi-classical analysis provides a coarse-grained explanation for the difference, the microscopic dynamics are fundamentally quantum mechanical.
%
This raises the question: is the asymptotic squeezing scaling also modified in the true quantum dynamics?

To answer this, we begin by developing a quantum interpretation of $\textrm{Var}[Y|Z]$.
Intuitively, $\textrm{Var}[Y|Z]$
can be computed as the remaining variance of $Y$ \emph{after} one has counter-rotated each $Z$-slice of the probability distribution
back to its original mean. 
Crucially, both semi-classically and quantum mechanically, 
this counter-rotation can be realized by evolving the system under one-axis twisting. 
From this perspective, the quantum analog of $\textrm{Var}[Y|Z]$ is simply:
\begin{equation}
     \textrm{Var}_\textrm{q} [Y|Z] = \bra{x} e^{it[H_{\rm XXZ}-\chi\frac{ \hat{Z}^2}{N}]} \; \hat{Y}^2 \; e^{-i t [H_{\rm XXZ} - \chi\frac{ \hat{Z}^2}{N}]} \ket{x}.  \label{eq:qvar}
\end{equation}
Since the thermalization induced by the XXZ-dynamics cannot be perfectly undone by one-axis twisting, $\textrm{Var}_\textrm{q} [Y|Z]$ will grow in time until saturating at its equilibrium value of $m_\textrm{xy}^2 N^2/2$.
In Fig.~\ref{fig:3}(e), we illustrate the variance growth for a 1D system at $\alpha=1.5, J_z=0$ and  find that 
 $\textrm{Var}_\textrm{q} [Y|Z]$ for $N=18$ indeed agrees extremely well with the semi-classical conditional variance in the thermodynamic limit [Fig.~\ref{fig:3}(e)].

A complementary explanation for the behavior of $\textrm{Var}_\textrm{q} [Y|Z]$ arises from the spectral structure of $H_\textrm{XXZ}$.
The ground states of adjacent magnetization sectors of $H_\textrm{XXZ}$ are connected  by the total spin raising and lower operator. 
Taken together, they form a so-called ``Anderson tower'' with energies scaling as $E \sim m^2/N$ [Fig.~\ref{fig:3}(f)].
In the one-axis-twisting model, squeezing arises from the fact that the \emph{entire} spectrum exhibits this ``Anderson tower'' structure. 
By contrast, at finite energy densities, the spectrum of $H_\textrm{XXZ}$ features only approximate ``Anderson towers": the raising and lowering  operator connect a given eigenstate in an $m$-sector to several others in  adjacent sectors, leading to a distribution of energies [Fig.~\ref{fig:3}(f)]. 
It is precisely the variance of this distribution which drives thermalization in the XXZ model, and thus the growth of $\textrm{Var}_\textrm{q}[Y|Z]$.

\emph{Outlook}.---Our work opens the door to a number of intriguing directions.
First, while we have established the presence of a phase transition between scalable squeezing and non-squeezing, the nature of this transition remains an open question. 
In particular, it would be interesting to derive the critical properties of the optimal squeezing solely from the universal properties of the CSB transition.
%
%
Second, our framework connecting the quantum Fisher information of pure states to spontaneous symmetry breaking suggests a new strategy for preparing metrologically useful states.  
However, the ability to prepare a state with large QFI does not immediately imply that one can straightforwardly utilize it for sensing; indeed, in the most general case, it is necessary to  time-reverse the dynamics, in order to extract the metrological signal~\cite{macri_loschmidt_2016, davis_approaching_2016, colombo_time-reversal-based_2022}.
In the case of $U(1)$ symmetry breaking, the emergence of squeezing \emph{does} guarantee a simple way to harness the metrological gain~\cite{wineland_squeezed_1994}.
Thus, generalizing our results to finite-temperature \emph{non-abelian} continuous symmetry breaking raises the question: is there any higher-symmetry analogue of spin-squeezing, which might enable sensitivity to perturbations beyond scalar fields \cite{forthcoming}?
%
%
%
Finally, our work implies that scalable squeezing can be realized in a variety of quantum simulation platforms with resonant dipolar interactions, including ultracold polar molecules, Rydberg atoms in tweezer arrays, and optically spin-defects in 2D quantum materials~\cite{ni2008polarmolecule,yan_observation_2013, de_leseleuc_optical_2017, chen_continuous_2022, chomaz_long-lived_2019, davis_probing_2021,tran_quantum_2016, gottscholl_initialization_2020, chejanovsky_single-spin_2021,eichhorn_optimizing_2019, smith_colour_2019}. 

\emph{Acknowledgements}--We gratefully acknowledge the insights of Ehud Altman, Joel Moore, Mike Zaletel, Chris Laumann and Francisco Machado. This work was support by the Army Research Office via grant number W911NF-21-1-0262 and through the MURI program (grant number W911NF-20-1-0136), and by the U.S. Department of
Energy via BES grant No.~DE-SC0019241 and via the National Quantum Information Science Research Centers, Quantum Systems Accelerator (QSA).
 E.J.D. acknowledges support from the Miller Institute for Basic Research in Science.
M.B. acknowledges support through the Department of Defense (DoD) through the National Defense Science and Engineering Graduate (NDSEG) Fellowship Program. B.R. acknowledges support from a Harvard Quantum Initiative postdoctoral fellowship. S.C. acknowledges support from the National Science Foundation Graduate Research Fellowship under Grant No. DGE 2140743. L.P. acknowledges support from FP7/ERC Consolidator Grant QSIMCORR, No. 771891, and the Deutsche Forschungsgemeinschaft (DFG, German Research Foundation) under Germany's Excellence Strategy -- EXC-2111 -- 390814868, as well as the Munich Quantum Valley, which is supported by the Bavarian state government with funds from the Hightech Agenda Bayern Plus. 
N.Y.Y acknowledges support from a Simons Investigator award. 
MPS simulations make use of the TenPy and ITensor libraries \cite{hauschild_efficient_2018, itensor} and QMC simulations make use of the ALPSCore libraries~\cite{gaenko_updated_2017,wallerberger_updated_2018}.

\bibliography{main-4-arxiv.bbl}

\begin{thebibliography}{10}
\bibitem{fn3}
\bibinfo{note}{For $\alpha > d$, we can choose $J_\perp = 1$ and let the $N$ go
  to infinity and maintain finite energy density. For $\alpha < d$, to have a
  finite energy density we need to let $J_\perp$ go to $0$ as $N$ goes to
  infinity. The all-to-all model is realized when we take $\alpha \to 0$ and
  $J_\perp = 1/N$. Our analysis is restricted to $\alpha > d$.}
\end{thebibliography}

\end{document}


\title{Supplementary Online Material for ``Scalable Spin Squeezing from Finite Temperature Easy-plane Magnetism''}

\maketitle

\tableofcontents

\section{Finite-temperature order phase diagram}
\subsection{Numerical methods}

\subsubsection{$d=1$: Purified matrix product states}

\begin{figure}
	\centering
 \includegraphics[width=0.99\textwidth]{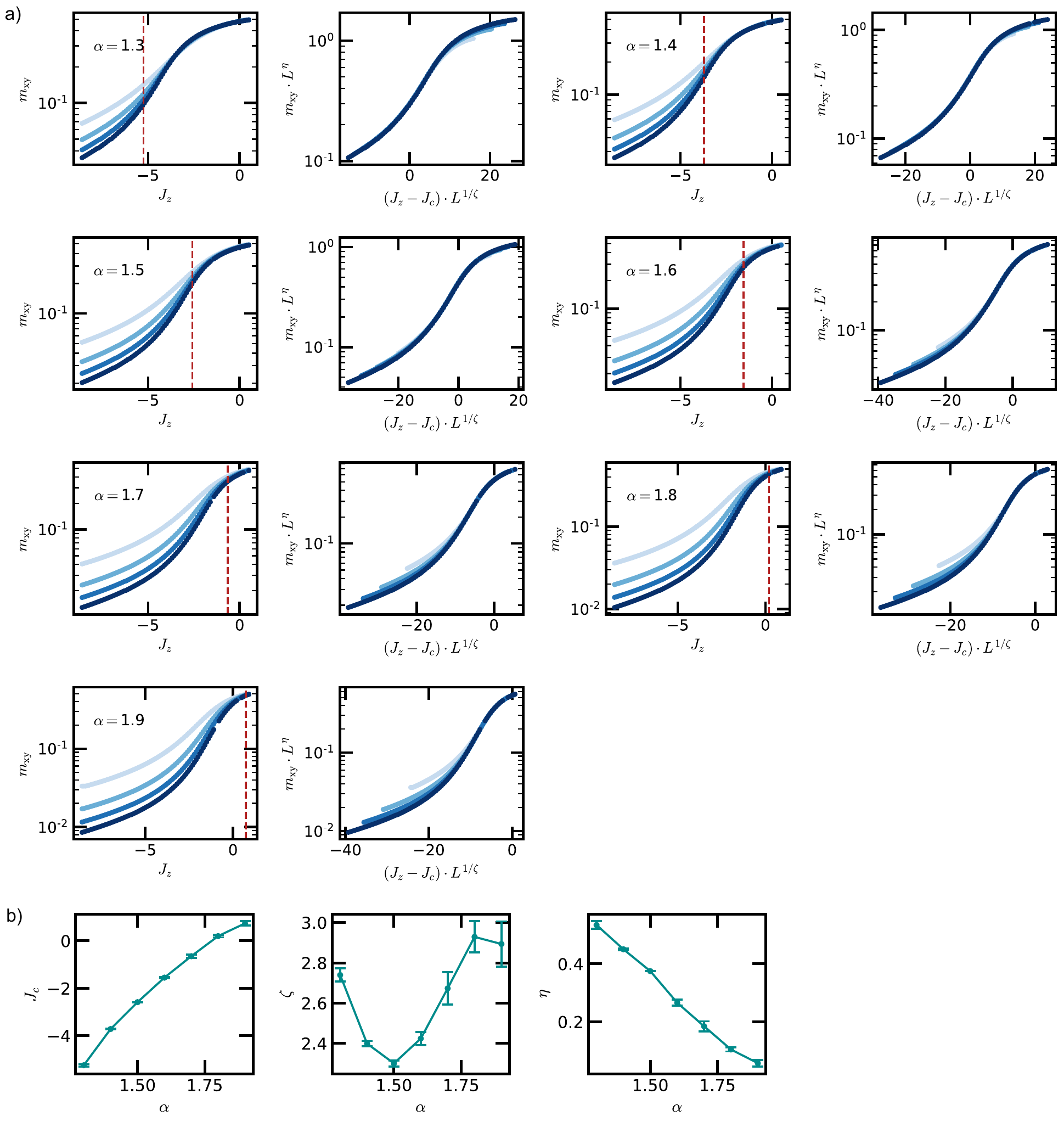}
	\caption{(a) Pairs of raw $m_{\rm xy}$ data (left) and collapsed data (right). The critical $J_c$ used for collapse is indicated by the dotted-red line. Increasing opacity indicates increasing system size, with $L=8,16,32,64$. (b) $J_c$, $\zeta$ and $\eta$ yielding the optimal collapse. The errors on $J_c$ are much smaller than analogous errors on $J_c^{\rm sqz}$ and are not relevant for our analysis.}
	\label{fig:mps}
\end{figure}

In $d=1$, we use imaginary time cooling of a purified matrix product state (MPS) to determine the equilibrium phase diagram. 
For each $\{\alpha, J_z\}$, we perform imaginary time evolution to cool an infinite temperature MPS to the effective temperature of $\ket{x}$ \cite{hauschild_efficient_2018}. 
Note that the cooling was performed with no limit on bond-dimension but with a truncation error threshold of $10^{-8}$. 
We then determine $m_{\rm xy}(\alpha, J_z)$ from the resulting density matrix and perform a finite size scaling analysis extract the critical Ising coupling strength, shown in Fig.~1(b). 
Specifically, we assume the magnetization to follow a typical scaling form 
\begin{equation}
    m_{\rm xy} = L^\eta f(\xi/L)
\end{equation}
where $\xi$ is the correlation length, assumed to diverge as $\sim (\frac{J_z-J_c}{J_c})^{-\zeta}$ (where we have used $\zeta$ as the correlation-length critical exponent -- as opposed to the conventional $\nu$ -- to avoid conflict with the notation in the main text). 
The finite-size collapse and associated critical exponents are shown in Fig.~\ref{fig:mps}. 
For a thermal phase transition in $d=1$, mean-field theory should apply for $\alpha \leq 3/2$.
We find exponents close the expected mean field values (e.g. $\zeta \approx 2.0$ at $\alpha=1.5$), but do not exactly recover the expected results.
We attribute this to the limited size used for our analysis ($L=8,16,32,64$) in combination with open boundary conditions and long-range interactions.

\subsubsection{$d=2$: Quantum Monte Carlo}
In $d=2$, the problem is more challenging and we rely on quantum Monte Carlo with worm-type updates \cite{Prokofev_worm_1998}. 
Since it is not possible to cool directly to the effective temperature of $\ket{x}$, we undertake a three parameter search over $\{\alpha, J_z, \beta\}$ (where $\beta$ is inverse temperature) to determine when the energy density at the critical temperature equals that of $\ket{x}$, yielding the the critical points in Fig.~1(c). 

For simplicity, we focus on the parameter range $3 \le \alpha < 4$ on the square lattice of linear size $L$ with periodic boundary conditions.
The model is sign-free for $J_{\perp} > 0$.  
We set $J_z = 1$ as the energy unit. The code is based on an adaptation of Ref.~\cite{Sadoune2022} and makes use of the ALPScore libraries~\cite{gaenko_updated_2017,wallerberger_updated_2018}.
Due to the long-range nature of the interactions the algorithm scales quadratically with system size. 
In practice we are therefore limited to linear system sizes $L \sim 30-40$.
We are interested in finding the critical temperature for spontaneous $U(1)$ symmetry breaking (easy-plane ferromagnetic order) as a function of $\{\alpha, J_z\}$.
If the corresponding energy is higher than the energy of the $x$-state, $ \ket{x} = \vert \rightarrow \cdots \rightarrow \rangle$ then, according to our conjecture, the system will exhibit scalable spin-squeezing. 
Our goal is therefore to find the maximum value of $J_z$ for a given $\alpha$ for which this is the case.

The critical behavior of long-range $U(1)$ models is a notoriously difficult problem~\cite{fisher_critical_1972,sak_recursion_1973,Luijten2001,horita_upper_2017,Angelini2014,Cescatti2019,giachetti_bkt_2021}.
For $\alpha \ge 4$ it is known that the long-range interactions are irrelevant and the model behaves in the same way as the nearest-neighbor model; i.e., true long-range order gives way to quasi long-range order characterized by power-law decay of the correlation functions~\cite{bruno2001absence}. 
In this case, the magnetization (or condensate density in the bosonic language)  vanishes in the thermodynamic limit for any nonzero temperature, but stiffness (or superfluid density in the bosonic language) remains finite.

For $\alpha < 4$ the generalized and inverse Mermin-Hohenberg-Wagner theorems~\cite{Burioni1999,Cassi1992} state that continuous symmetry breaking \emph{is allowed} at finite temperature, as is indeed observed in our simulations.
For the parameter range $3 \le \alpha < 4$, we apply the most widely accepted theory of criticality for long-range interactions, namely that of Sak~\cite{sak_recursion_1973}, which asserts that the transition is continuous and that the anomalous critical exponent, $\eta$, is unrenormalized and hence behaves as ${\rm max}(4 - \alpha, 0.25)$. 
From this formula one can appreciate the difficulty in determining the universality for $\alpha \ge 3.75$. We refer to Refs.~\cite{defenu_long-range_2021,giachetti_bkt_2021} for an in-depth discussion.

To the best of our knowledge, large-scale quantum Monte Carlo simulations for the $XY$-model on non-dilute lattices have never been published before. 
On diluted graphs however, spontaneous symmetry breaking has been observed numerically~\cite{Berganza2013, Cescatti2019}.
We note that while all our data are consistent with the Sak scenario (and we apply hence this theory, cf Ref.~\cite{Angelini2014}), we also observe that the corrections to scaling are particularly strong in the vicinity of $\alpha = 3.75$ (and possibly logarithmic in nature~\cite{Angelini2014,horita_upper_2017}). Such corrections cannot possibly be fully captured on the system sizes that can be studied by current quantum Monte Carlo simulations for the long-range XXZ-model.

As mentioned above, true long-range order can be distinguished from quasi-long range order by computing the condensate density, $n_0$, defined  as 
\begin{equation}
n_0 = \frac{1}{L^2} \sum_{\mathbf{j}} \mathcal{G}( \mathbf{j}, \tau=0),
\label{eq:def_condfrac}
\end{equation}
where $ \mathcal{G}( \mathbf{j}, \tau=0) = \left< \sigma^+_{\mathbf{j}} \sigma^-_{\mathbf{0}} + {\rm h.c.} \right>$ is the equal-time off-diagonal spin correlation function for a system with translational invariance.  
Note that we did not resum the interaction potential over periodic images. 
We found it more convenient not to do so when comparing spin correlation functions for different system sizes (because we approach the critical temperature from the side of slightly larger temperatures). 
Furthermore, within our resolution we saw no noticeable difference for the energy between the two approaches (and they should be equivalent in the thermodynamic limit).

As in the $d=1$ case, in the theory of second order phase transitions, universality classes are characterized by two independent critical exponents, $\zeta$ and $\eta$. The former expresses the divergence of the correlation length when approaching the critical point, the latter is the anomalous decay of the correlation function  $\mathcal{G}$ at the critical point.
The theory of finite size scaling predicts for thermal transitions the following behavior for the singular part of the condensate fraction,
%
%
\begin{equation}
n_{0,s}(\epsilon, L) \sim L^{-\eta} Y_{n_0}(\xi / L),
\label{eq:FSS_n0}
\end{equation}
where $\xi = \epsilon^{-\zeta}$ is the correlation length,  $\epsilon$ is the dimensionless detuning from the the inverse critical temperature, $\epsilon = \frac{\beta - \beta_c}{\beta_c}$, and $ Y_{n_0}$ is an unknown, universal function. 
If there is spontaneous symmetry breaking, then the curves $n_0(\beta)L^{\eta}$ intersect for various $L$ in a single point corresponding to the inverse critical temperature, up to scaling corrections.
In this manner, we can determine $\beta_c(\alpha, J_z)$, as well as the associated energy density, and thereby derive the equilibrium phase diagram shown in Fig.~1(c).

\begin{figure}
	\centering
 \includegraphics[width=0.99\textwidth]{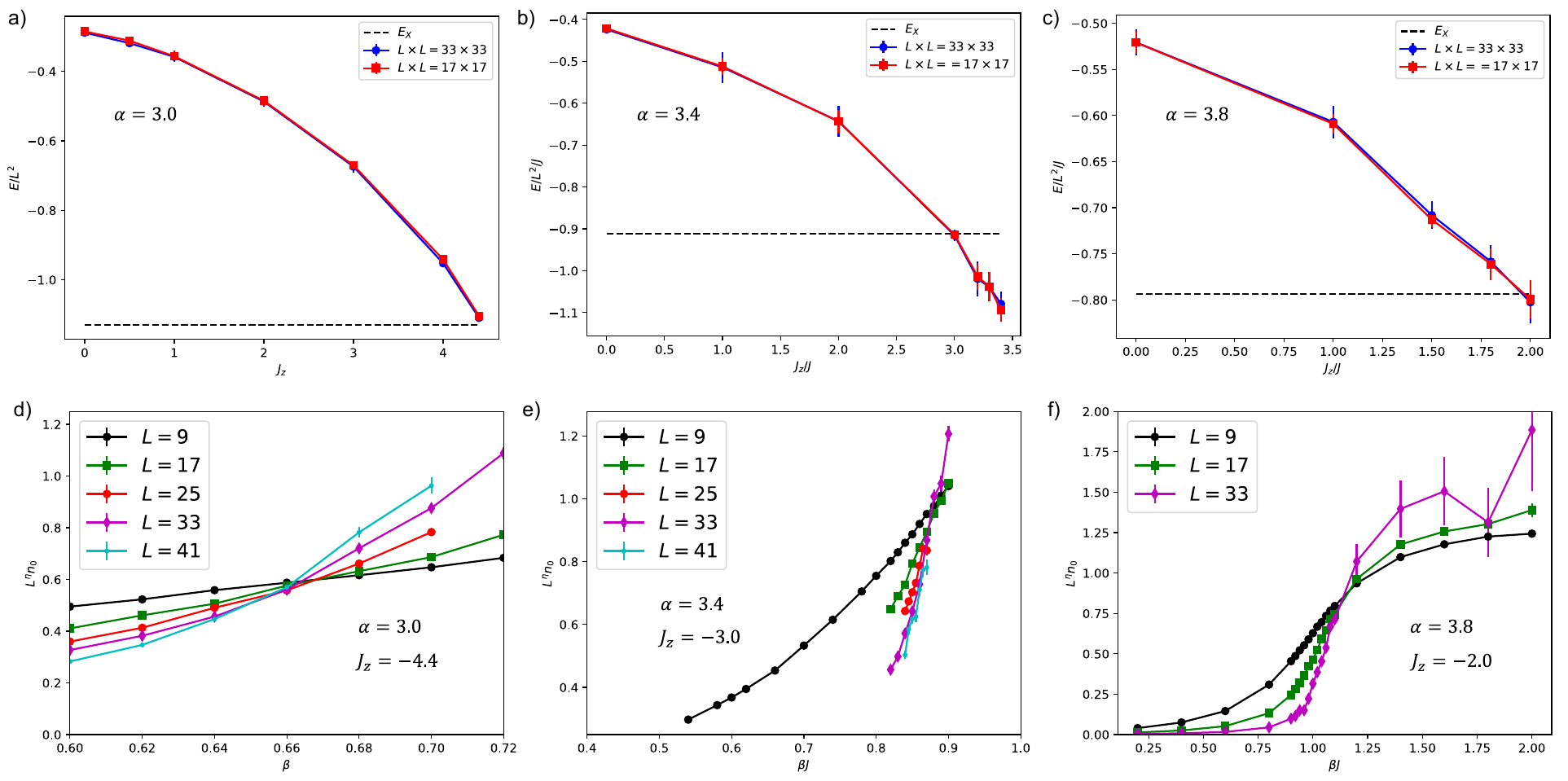}
	\caption{(a-c) The energy density at $\beta_c$ as a function of $J_z$ for $\alpha=3.0, 3.4, 3.8$ respectively. Results for $\alpha=3.2, 3.6$ are similar. The $\ket{x}$ energy density, indicated by the dashed-horizontal line, eventually intersects the critical energy density as $J_z$ increases, yielding $J_c$ shown in Fig.~1(c). (d-f) Examples of the condensate-fraction finite-size scaling analysis for $\alpha=3.0,3.4,3.8$ at the associated $J_c$. Results for $\alpha=3.2,3.6$ are similar. The sharp crossing enables accurate determination of $\beta_c$.}
	\label{fig:qmc}
\end{figure}

Fig.~\ref{fig:qmc} shows example results of this procedure.
In general, we see that the energy density at $\beta_c$ decreases with increasing $J_z$ as expected Fig.~\ref{fig:qmc}(a-c), eventually intersecting the energy density of $\ket{x}$ (indicated by the horizontal line) at $J_c$.
As in the $d=1$ case, the errors on the equilibrium $J_c$ are much smaller than on $J_c^{\rm sqz}$, and we neglect them.
We also note that we observe excellent crossings for the condensate fraction [Fig.~\ref{fig:qmc}(d-f)], showing that we can identify $\beta_c$ very accurately (and therefore the associated energy density).

\subsection{Analytic methods}
In addition to these numerical approaches, we develop an analytic approximation for the finite-temperature symmetry breaking transition.
%
The key issue we wish to resolve analytically is whether order, at the effective temperature of the coherent-spin state (CSS) $\ket{x}$, requires $\alpha<2$ for $J_z<1$ (as opposed to persisting at $\alpha=2$). 
To this end, we focus on the vicinity of $\alpha \lesssim 2d$, $J_z \lesssim 1$ where the system can be modeled as Bose gas, based on  previous work in Refs.~\cite{nakano1994quantum,nakano1994quantum2}, which were in turn motivated by the exact solution of the Haldane--Shastry spin chain \cite{haldane1988exact,haldane1991spinon}.
The key physical intution is that the ground state manifold of the model with $SU(2)$ symmetry contains $\ket{x}$, and with weak anisotropy, $0 \leq \delta \equiv 1-J_z << 1$, this state remains at a low effective temperature so the relevant equilibrium states are still well described as Gaussian states with few excitations.
%

%

\subsubsection{Holstein--Primakoff bosonization: using z-vacuum}

We perform a Holstein--Primakoff bosonization of the model, assuming the fully polarized state along z-axis as the vacuum and making a large-$S$ approximation:
\begin{equation}
\begin{gathered}
S^+_i = ({2S-b^\dag_i b_i})^{1/2}\;b_i~, \\
S^-_i = b^\dag_i ({2S-b^\dag_i b_i})^{1/2}~, \\
S^z_i = S - b^\dag_i b_i~,
\end{gathered}
\end{equation}
where $\sqrt{2S-b^\dag_i b_i} = \sqrt{2S}\left(1 - \frac{1}{4S}b^\dag_i b_i + O(1/S)\right)$.
This approach may seem surprising, however we observe that it is necessary to choose a vacuum that respects $U(1)$ symmetry in order to capture the ordering transition.
%
Consequently, the vacuum has finite energy density except at the spin-isotropic point, where it is a member of the degenerate ground state manifold.
The applicability of this theory away from this point is nontrivial: however, as we shall see, the primary effect of anisotropy is to introduce an energy offset between symmetry sectors, and it does not affect the single-particle dynamics at leading order. 

Fourier transforming via $b^\dag_i = \frac{1}{\sqrt N} \sum_{\bm q} e^{-i\bm q \cdot \bm r_i} b^\dag_{\bm q}$ 
and defining $\eta(\bm q) = \sum_{\bm r \neq 0} |\bm r|^{-\alpha} e^{i\bm q \cdot \bm r}$, the power-law XXZ Hamiltonian is represented in momentum space as
\begin{equation}
\begin{gathered}
H = S^2 H^{(0)} + S H^{(2)} + H^{(4)} + O(1/S)~, \\
H^{(0)} = {\rm constant}~, \\
H^{(2)} = \sum_{\bm q} b^\dag_{\bm q} b_{\bm q} (\omega(\bm q)-\delta\eta(0))~,\quad\omega(\bm q) \equiv \eta(0) - \eta(\bm q)~, \\
H^{(4)} = \frac{1}{8N} \sum_{\substack{\bm q,\bm q',\\\bm q'',\bm q'''}} b^\dag_{\bm q} b^\dag_{\bm q'} b_{\bm q''} b_{\bm q'''} \delta(\bm q+\bm q'-\bm q''-\bm q''') \left(\eta(\bm q) + \eta(\bm q') + \eta(\bm q'') + \eta(\bm q''') - 4 J_z \eta(\bm q-\bm q'')\right)~.
\end{gathered}
\end{equation}
As $\bm q$ is a momentum eigenvalue, $\eta(\bm q) = \eta(-\bm q) = \eta(\bm q)^\ast$; so $\eta(\bm q)$ is real.
We proceed following a standard treatment by minimizing the free energy in the variational manifold of Gaussian states of the form $\ket{\{n_{\bm q}\}} = \prod_{\bm q} (n_{\bm q}!)^{-1/2} (b^\dag_{\bm q})^{n_{\bm q}}\ket{\Psi_\text{vac}}$, keeping only terms up to $O(1/S)$. 
The minimization depends only on the effective single particle dispersion $\e(\bm q) = \frac{\partial \langle{H}\rangle}{\partial n_{\bm q}}\big|_{n_{\bm q' \neq \bm q}}$, and quite generally obtains the Bose-Einstein distribution \cite{reif}.
%
In the disordered phase this results in the self-consistency condition
\begin{equation}
M = \sum_{\bm q} n(\bm q) = \sum_{\bm q} \frac{1}{e^{(\e(\bm q)-\mu)/T}-1}.
\label{eq:bose_scc}
\end{equation}
We emphasize that Eq.~\eqref{eq:bose_scc} only counts particles in \emph{excited} modes due to the vanishing density of states at $q=0$.
Therefore, while the variational states have fixed particle number $M$ we are able to treat the problem in the \emph{grand canonical} ensemble by allowing the Bose-Einstein condensate to act as a source of particles, whose average is set by $\mu$.
Above $T_c$ one can self-consistently determine $\mu<0$, and we identify $T_c$ as the temperature satisfying Eq.~\eqref{eq:bose_scc} with $\mu=0$, where exactly $M$ particles are extracted from the condensate into excited modes.

To determine $\e(\bm q)$ explicitly, we apply Wick's theorem and Eq.~\eqref{eq:bose_scc} to find
\begin{align}
\langle{H}\rangle &= \eta(0) \left(-\frac12 N S^2 J_z - \delta \left( S M -\frac{M(M-1)}{2N}\right)\right) + \sum_{\bm q} n(\bm q) \left(S-\frac{M-1}{2N}\right) \omega(\bm q) \nonumber \\
&\hspace{2cm} - \frac{1}{2N} \sum_{\bm q} n(\bm q)  \sum_{\bm q' \neq \bm q} n(\bm q') (J_z \eta(\bm q-\bm q') - \eta(\bm q'))~.
\label{eq:E_quad}
\end{align}

From Eq.~\eqref{eq:E_quad}, we can gain further insight into the validity of our approximation away from the Heisenberg point.
Specifically, while the anisotropy $\delta$ contributes an $M$-dependent offset leading to a unique ground state sector with $M=SN$, it does not directly modify the quadratic terms.
While $\ket{x}$ superposes particle number sectors, the fluctuations are small and we work in the sector $M=\langle{M}\rangle=SN$.
This renders the leading order effect of the anisotropy irrelevant.
Moreover, for small anisotropy, which perturbs $\ket{x}$ away from the ground-state manifold, interactions are suppressed, as for low temperature only very low momentum modes have significant occupation. 
The quartic term is then suppressed by the width of the momentum distribution $n(\bm q)$.
%


\subsubsection{Condensation and effective initial temperatures}

Based on the above, as an approximation, we discard the interaction term, making $H$ diagonal in $\bm q$.
Now there is no remaining $\bm q$-dependent term involving anisotropy, so this result is the same as that of Refs.~\cite{nakano1994quantum,nakano1994quantum2} for the $SU(2)$ model.

To leading order, the estimate of the critical temperature is
\begin{equation}
T_c = \begin{cases}
-\dfrac{\pi S}{2 \Gamma(\alpha) \cos(\frac{\pi \alpha}{2})}\left(\dfrac{\pi S (\alpha-1)}{ \Gamma(\frac{1}{\alpha-1}) \zeta(\frac{1}{\alpha-1})}\right)^{\alpha-1}~,&d=1~,\\
-\dfrac{2^{1-\alpha} \pi^2 S}{\Gamma(\frac\alpha2)^2 \sin(\frac{\pi \alpha}{2})}\left(\dfrac{2 \pi S (\alpha-2)}{ \Gamma(\frac{2}{\alpha-2}) \zeta(\frac{2}{\alpha-2})}\right)^\frac{\alpha-2}{2}~,&d=2~.
\end{cases}
\label{eq:tc}
\end{equation}
In $d=1$ the leading behavior as $\alpha\to2$ from below is $T_c \sim \frac{\pi^2}{2} S^2 (2-\alpha)$.
In $d=2$ the critical temperature jumps discontinuously at $\alpha=4$, from $\lim_{\alpha\to4} T_c = \frac{\pi^2}{8}$ to 0, as required by rigorous bounds \cite{bruno2001absence}.

Using the same picture of the low-energy thermodynamics allows to compute the effective temperature $T_0$ of the CSS initial state.
Its energy is exactly $E_\mathrm{CSS} = -\frac{N}{8} \eta(0)$, which turns out to be (as $N\to\infty$) the lowest variational energy for $M=SN=\frac N2$.
Given the ground state energy density, we can compute the excitation energy per site of the CSS and relate this to its temperature in the Bose gas.
To leading order
\begin{equation}
T_0 = \begin{cases}
\left(-\dfrac{\pi S}{2 \Gamma(\alpha) \cos(\frac{\pi \alpha}{2})}\right)^\frac1\alpha \left(\dfrac{\pi (\alpha-1) \mathcal E(\alpha,J_z)}{ \Gamma(\frac{\alpha}{\alpha-1}) \zeta(\frac{\alpha}{\alpha-1})}\right)^\frac{\alpha-1}{\alpha}~,& d=1~,\\
\left(-\dfrac{2^{1-\alpha} \pi^2 S}{\Gamma(\frac\alpha2)^2 \sin(\frac{\pi \alpha}{2})}\right)^\frac2\alpha \left(\dfrac{2 \pi (\alpha-2) \mathcal E(\alpha,J_z)}{\Gamma(\frac{\alpha}{\alpha-2})\zeta(\frac{\alpha}{\alpha-2})}\right)^\frac{\alpha-2}{\alpha}~,&d=2~,
\end{cases}
\label{eq:tcss}
\end{equation}
where $\mathcal E(\alpha,J_z)$ is the energy density of the CSS in the thermodynamic limit.

\begin{figure}[h!]
	\centering
 \includegraphics[width=0.65\textwidth]{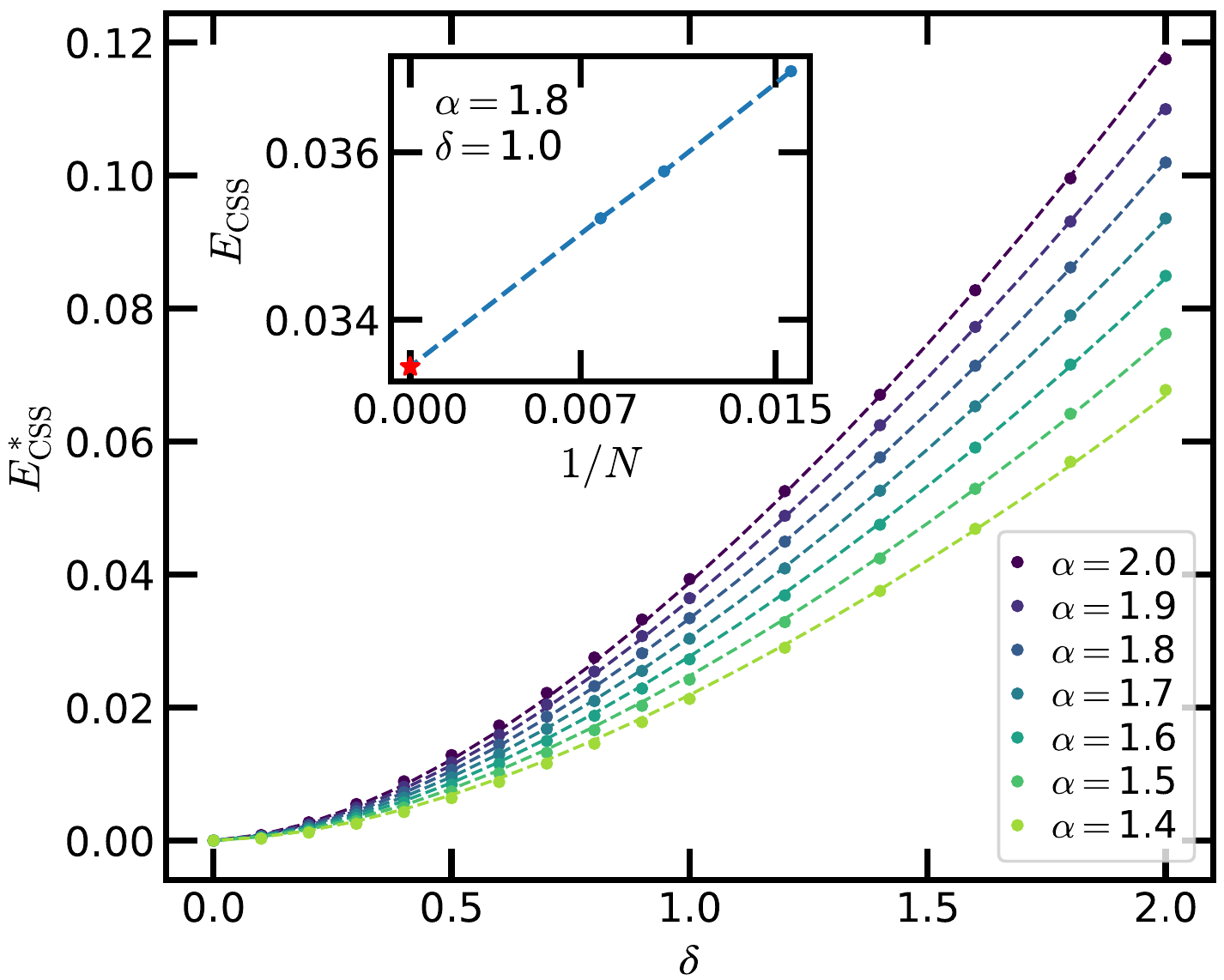}
	\caption{Interpolations of $E^*_{\rm CSS}$ used to determine phase boundary as described in methods. (inset) Example extrapolation of DMRG results to determine $E^*_{\rm CSS} \equiv \lim_{N \to \infty} E^N_{\rm CSS}$.}
	\label{fig:5}
\end{figure}

To determine $\mathcal{E}(\alpha,J_z)$, we extrapolate results from DMRG on periodic systems of length $N=64,96,128$ using ITensor \cite{itensor}~(Fig.~\ref{fig:5} inset).
This provides an unbiased estimate accounting for all quantum fluctuations, at the cost of only being able to compute $T_0$ for specific points in the parameter space.
To estimate the critical $J_c(\alpha)$ such that $T_0 = T_c$ more precisely, we we use a simple polynomial fit of $\mathcal E(\alpha,J_z)$~(Fig.~\ref{fig:5}).
The resulting phase boundary is shown in gold in Fig.~1(b) of the main text, and shows good agreement with the boundary obtained from MPS numerics, especially as $J_z \rightarrow 1$ where this approximation should be most accurate. 

\section{Extraction of the effective one-axis-twisting strength}
\begin{figure}[h!]
	\centering
 \includegraphics[width=0.8\textwidth]{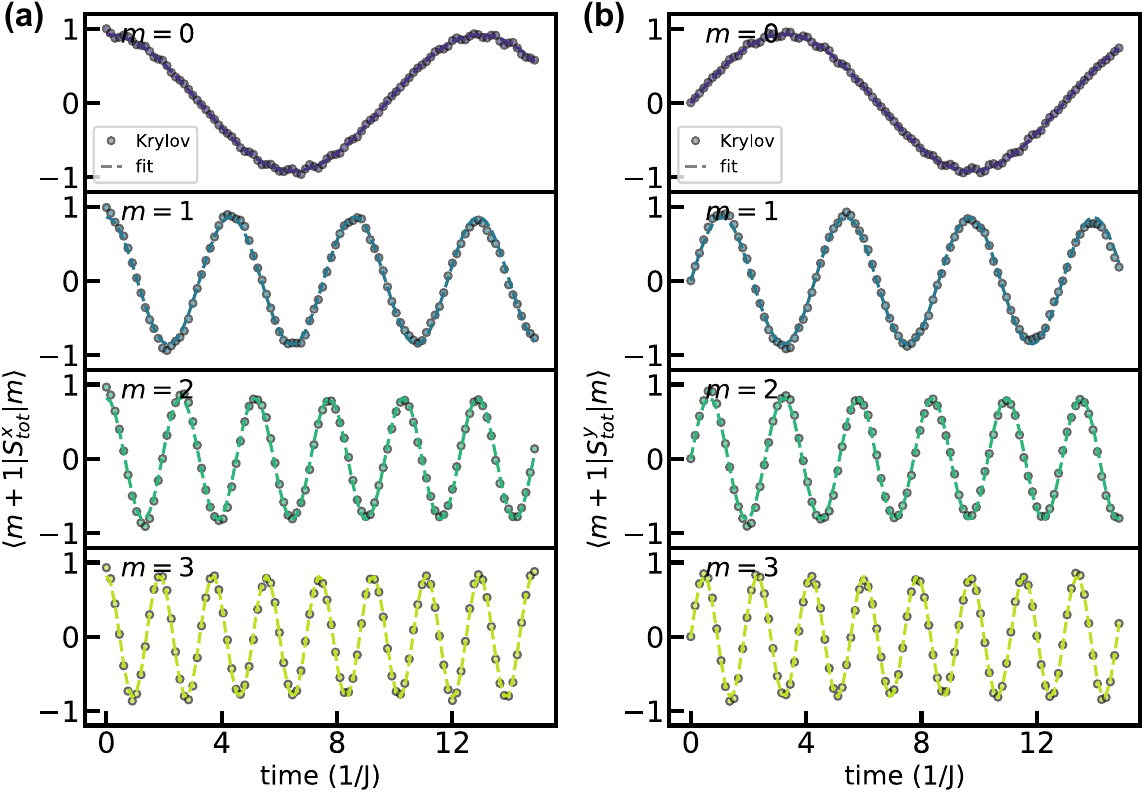}
	\caption{(a,b) Time-dependent matrix elements of $S^x_{\rm tot}$, $S^y_{\rm tot}$ respectively used to determine $\Delta E$ as described in the Methods. The $\Delta E$ are then used to determine $\chi$ from $\Delta E = \chi(2m+1)$.}
	\label{fig:4}
\end{figure}

To compute the quantum analog of the conditional variance $\mathrm{Var_q}[Y|Z]$ from the Loschmidt echo (Eq.~3 of the main text), we need to determine the effective OAT strength $\chi$. 
Semiclassically, $\chi$ is defined so that different $m$-sectors are rotated back to their original positions. 
Quantum mechanically, this corresponds to cancelling the relative phase accumulation between two adjacent $m$-sectors, i.e. we require:
\begin{equation}
    \langle m+1| e^{it(H_{\rm XXZ}-\chi \frac{\hat{Z}^2}{N})} \; \hat{Y} \; e^{-i t (H_{\rm XXZ} - \chi  \frac{\hat{Z}^2}{N})} |m\rangle = 0,
\end{equation}
where $|m\rangle$ is the projection of $|x\rangle$ into the $m$-sector. 
This immediately leads to the expectation:
\begin{equation}
\begin{split}
    \langle m+1| e^{itH_{\rm XXZ}} \; \hat{X} \; e^{-i t H_{\rm XXZ}} |m\rangle &\sim \mathrm{cos}(\Delta E \cdot t),\\
    \langle m+1| e^{itH_{\rm XXZ}} \; \hat{Y} \; e^{-i t H_{\rm XXZ}} |m\rangle &\sim \mathrm{sin}(\Delta E \cdot t),
\end{split}
\label{Eq:oscillation_exact_chi}
\end{equation}
where $\Delta E$ is the relative energy difference between two adjacent $m$-sectors.
To determine $\Delta E$, we compute the left side of Eq.~\ref{Eq:oscillation_exact_chi} by time evolving $|m\rangle$ and $|m+1\rangle$ with the Krylov subspace method and then fit the data to sinusoid oscillations~(Fig.~\ref{fig:4}).
We note these oscillations are damped in principle (reflecting the linear growth of $\mathrm{Var_q}[Y|Z]$), but this is neglible in our fits.
We then extract $\chi$ by fitting $\Delta E = (2m+1)\chi$.

\section{Heisenberg Limited Sensing With Non-interacting Bosons}
To provide a very simple and concrete example of how long-range order leads to quantum enhanced sensitivity, we consider a toy model of hopping bosons at zero temperature.
\begin{equation}
    H = -\sum_{\langle i j \rangle} a^\dagger_i a_j + h.c.
\end{equation}
For this quadratic Hamiltonian of $N$ hopping bosons, the ground state indeed exhibits off-diagonal long-range order and is given by the Fock state with $N$ particles in the zero-momentum mode. Let us define the creation operator associated with the zero-momentum  mode as,
\begin{equation}
    \bar{a}^\dagger \equiv \frac{1}{\sqrt{N_{\rm sys}}} \sum_i a_i^\dagger,
\end{equation}
where $N_{\rm sys}$ is the number of sites in the system. 
Assuming that one is interested in measuring a uniform external electric field, $\mathcal{E}_\textrm{ext} \sim E_0 \sum_i (a_i +a_i^\dagger)$, the relevant variance that determines the sensitivity of the zero-temperature state is given by,
\begin{align}
    \bra{N}_{k=0} \sum_{i}(a_i + a^\dagger_i)^2 \ket{N}_{k=0} &= \bra{N}_{k=0} \sum_{i,j}(2a_i^\dagger a_j+N_{\rm sys}) \ket{N}_{k=0} \\
    &= N_{\rm sys} \bra{N}_{k=0} 2\bar{a}^\dagger \bar{a}+1 \ket{N}_{k=0} \\
    &= N_{\rm sys}(2N+1)
\end{align}
Provided $N \propto N_{\rm sys}$, we obtain a variance that scales quadratically with $N$ and thus a Heisenberg limited sensitivity
It is crucial to note that making use of this Heisenberg limited sensitivity requires the ability to perform a projective measurement onto the initial state of the system -- in this case, the $N$ particle zero-momentum Fock state. 

\section{Absence of large QFI in the Ferromagnetic Heisenberg Model}

In the main text, we have made the statement that an order parameter off-diagonal in the symmetry sectors in systems with continuous symmetry breaking is important for the development of large QFI. 
Here, we provide an example to illustrate this point. 
Let us consider the $3D$ ferromagnetic Heisenberg model: 
\begin{equation}
    H=-J\sum_{\langle i j \rangle} \vec{S}_i \cdot \vec{S}_j
    \label{eqn1}
\end{equation}
where the sum is over nearest neighbor interactions.
On the one hand, this model \emph{does} have finite-temperature $SU(2)$ breaking order.
However, it also has an obvious problem for our protocol: the lowest-temperature product states, i.e.~maximally polarized spins, are eigenstates, and trivially never develop long-range connected correlations.
Even introducing some disorder into the state so that it is no longer an eigenstate does not fix the problem --- instead, the state will simply develop an $O(1)$ correlation length as the local spin direction relaxes to a uniform value. 
The issue of course is that the order parameter $\hat{n} \cdot \vec{S}$ is itself a symmetry of $H$, violating condition 1.
To overcome this problem, we simply need a model that breaks $\mathrm{SU}(2)$ with an order-parameter that connects total-spin sectors.
Indeed, an \emph{anti-ferromagetic} Heisenberg model fully satisfies the above requirements and \emph{does} evolve large QFI in the quench dynamics of a N\'eel state \cite{forthcoming}.

\section{Short-time non-scalable squeezing}
As mentioned above, apart from scalable squeezing at late times there can also be short-time \emph{non-scalable} squeezing.
We now explain the origin of this short-time squeezing using spin-wave theory and thereby show it is indeed non-scalable and should not affect the squeezing phase diagram.

\subsection{Holstein--Primakoff bosonization: using x-vacuum}
To setup the spin-wave analysis, we perform a Holstein--Primakoff bosonization of the spin-$\frac{1}{2}$ degrees of freedom defining $\ket{x}$ to be the vacuum:
\begin{equation}
\begin{split}
    \sigma^x_i &= \frac{1}{2}-b^\dag_i b_i,\\
    \sigma^y_i &=\mathrm{i} \left(({1-b_i^\dag b_i})^{1/2}\;b_i - b^\dag_i({1-b_i^\dag b_i})^{1/2}\right),\\
    \sigma^z_i &= \left(({1-b_i^\dag b_i})^{1/2}\;b_i + b^\dag_i({1-b_i^\dag b_i})^{1/2}\right),
\end{split}
\label{eq:HP_repre}
\end{equation}
where $b_i$ ($b^\dag_i$) is the annihilation (creation) operator for the HP boson on site $i$. 
Plugging these definitions into Eq.~1 in the main text, keeping only the quadratic terms, and Fourier transforming so $b^\dag_i = \frac{1}{\sqrt N} \sum_{\bm q} e^{-i\bm q \cdot \bm r_i} b^\dag_{\bm q}$, we obtain
\begin{equation}
\begin{split}
    H_\mathrm{quad} = -\frac{\Delta-1}{4}&\eta_0(2b^\dag_0 b_0+b^\dag_0 b^\dag_0+b_0 b_0)\\
    &-\sum_{{\bm q}\neq 0} (\frac{\Delta+1}{2}\eta_{\bm q}-\eta_0)b^\dag_{\bm q} b^{}_{\bm q}+\frac{\Delta-1}{4}\eta_{\bm q} (b^\dag_{\bm q} b^\dag_{-{\bm q}}+b_{\bm q} b_{-{\bm q}}), 
\end{split}
\end{equation}
where $\Delta \equiv J_z / J_\perp$, we have set $J_\perp=1$, and $\eta_{\bm q} = \sum_{\bm r \neq 0} |\bm r|^{-\alpha} \mathrm{e}^{i\bm q \cdot \bm r}$ is the Fourier transform of the long-range interaction. 

\subsection{Early-time squeezing}
We first note that the the ${\bm q}=0$ term is nothing but a bosonic analog of the OAT model -- that is, $\sigma^z_{q=0} \approx b_0^\dag + b_0$.
This term \emph{may} generate late-time scalable squeezing depending on how it is modified by higher-order terms, a point we will revisit in detail below.
At short times, however, the ${\bm q}\neq 0$ modes also contribute to the squeezing dynamics.
In particular, we calculate their contribution to the variance of total spin with an angle $\theta$ in the $yz$-plane:
\begin{equation}
\begin{split}
    \langle (\sigma^\theta_{tot})^2\rangle &= 2\cos(2\theta)\sum_{{\bm q}\neq 0}\mathrm{Re}[\langle b_{\bm q} b_{-{\bm q}} \rangle(t)]+2\sin(2\theta)\sum_{{\bm q}\neq 0}\mathrm{Im}[\langle b_{\bm q} b_{-{\bm q}} \rangle(t)] \\
& \hspace{10mm} + ({\rm contribution ~ from}~{\bm q}=0)+(\theta~{\rm independent}~{\rm terms}).
\end{split}
\label{eq:min_variance}
\end{equation}
By diagonalizing $H_\mathrm{quad}$ using a Bogoliubov transformation, we obtain
\begin{equation}
    \langle b_{\bm q} b_{-{\bm q}} \rangle(t) =  -\frac{A_{\bm q} B_{\bm q}}{\omega_{\bm q}^2}\sin^2(\omega_{\bm q} t)+i \frac{B_{\bm q}}{2\omega_{\bm q}}\sin(2\omega_{\bm q} t), 
\label{eq:pmode_solution}
\end{equation}
where $A_{\bm q} = \frac{\Delta+1}{2}\eta_{\bm q}-\eta_0$, $B_{\bm q}=\frac{\Delta-1}{2}\eta_{\bm q}$, and $\omega_{\bm q} = \sqrt{A_{\bm q}^2-B_{\bm q}^2}$. 
Combining the preceding equations, we find that the ${\bm q}\neq 0$ modes contributes an extra reduction to the minimum value of $\langle (\sigma^\theta_{tot})^2\rangle$, by an amount of at most
\begin{equation}
-2\sqrt{[\sum_{{\bm q}\neq 0}\frac{A_{\bm q} B_{\bm q}}{\omega_{\bm q}^2}\sin^2(\omega_{\bm q} t)]^2+[\sum_{k\neq 0}\frac{B_{\bm q}}{2\omega_{\bm q}}\sin(2\omega_{\bm q} t)]^2}. 
\label{eq:reduction_variance}
\end{equation}
This reduction behaves as a damped oscillation, which leads to a few local minima in the variance and thus the squeezing parameter at short times. 

Two remarks are in order. 
First, the extra reduction in variance (Eq.~\ref{eq:reduction_variance}) converges in the thermodynamic limit by replacing the summation by $N$ times an integral, showing that the early-time squeezing is essentially a local process and does \emph{not} scale with system size.
This justifies the above data analysis which excludes the early-time minima when determining the squeezing phase diagram. 

Second, the ${\bm q}=0$ term in the $H_\mathrm{quad}$ does \emph{not} automatically entail late-time scalable squeezing.
There is a modification to the effective squeezing strength due to the higher-order terms.
In particular, at quartic order, we have additional terms:
\begin{equation}
\begin{split}
    H_\mathrm{aniso} &=\frac{\Delta-1}{8N}\sum_{{\bm q}+{\bm q'}={\bm q''}+{\bm q'''}}(\eta_{{\bm q''}}+\eta_{{\bm q'''}})b^\dag_{{\bm q}} b_{-{\bm q'}}(b_{{\bm q''}} b_{{\bm q'''}}+b^\dag_{-{\bm q''}} b^\dag_{-{\bm q'''}}+b^\dag_{-{\bm q''}} b^{}_{{\bm q'''}}+b^{}_{\bm q''} b^\dag_{-{\bm q'''}}),\\
    H_{\mathrm{iso}}
    &=\frac{1}{4N}\sum_{{\bm q}+{\bm q'}={\bm q''}+{\bm q'''}}(2\eta_{{\bm q}-{\bm q''}}+2\eta_{{\bm q}-{\bm q'''}}-\eta_{{\bm q}}-\eta_{{\bm q'}}-\eta_{{\bm q''}}-\eta_{{\bm q'''}})b^\dag_{\bm q} b^\dag_{\bm q'}b^{}_{\bm q''} b^{}_{\bm q'''}. 
\end{split}
\end{equation}
Within a mean-field treatment, we replace the ${\bm q}\neq 0$ modes with their late-time equilibrium values, and extra quadratic terms for the ${\bm q}=0$ mode emerge as:
\begin{equation}
\begin{split}
    H_{\mathrm{aniso}} &\approx \frac{\Delta-1}{8}\sum_{{\bm q}}(\eta_{{\bm q}}+\eta_{0})\left[\begin{split}\left(\langle b^\dag_{\bm q} b_{\bm q}\rangle +\mathrm{Re}\langle b_{\bm q} b_{-{\bm q}}\rangle\right)&(2b^\dag_{0} b^{}_{0}+b^{}_{0} b^{}_{0}+b^\dag_{0} b^\dag_{0})\\&+\\
    \mathrm{i}~\mathrm{Im}\langle b_{\bm q} b_{-{\bm q}}\rangle&(b^\dag_{0} b^\dag_{0}-b^{}_{0} b^{}_{0})\end{split}\right]
    \\
    H_{\mathrm{iso}}& \approx \sum_{{\bm q}}(\eta_{{\bm q}}-\eta_{0})
   \left[ \mathrm{Re}\langle b_{\bm q} b_{-{\bm q}}\rangle(b^{}_{0} b^{}_{0}+b^\dag_{0} b^\dag_{0})+\mathrm{i}~\mathrm{Im}\langle b_{\bm q} b_{-{\bm q}}\rangle(b^\dag_{0} b^\dag_{0}-b^{}_{0} b^{}_{0}) \right].
\end{split}
\end{equation}

By generalizing this analysis to sufficiently high order in the expansion, it may be possible to determine the squeezing phase diagram analytically based on when the ${\bm q}\neq 0$ term vanishes.
This is an interesting direction for future research.


\section{Benchmarking the Discrete Truncated Wigner Approximation}

The strongest numerical evidence supporting our conjecture derives from large-scale discrete truncated Wigner approximation (DTWA) simulations \cite{schachenmayer}.
Although it is not possible to directly check the results of DTWA at the relevant system sizes, we  can compare it to Krylov-based time evolution at small system sizes ($N \sim 15$) and time-dependent variational Monte Carlo (tVMC) at at moderate system sizes ($N \sim 100$).
We note that tVMC has been shown to give accurate results for squeezing in \cite{comparin}.

\begin{figure}
	\centering
 \includegraphics[width=0.99\textwidth]{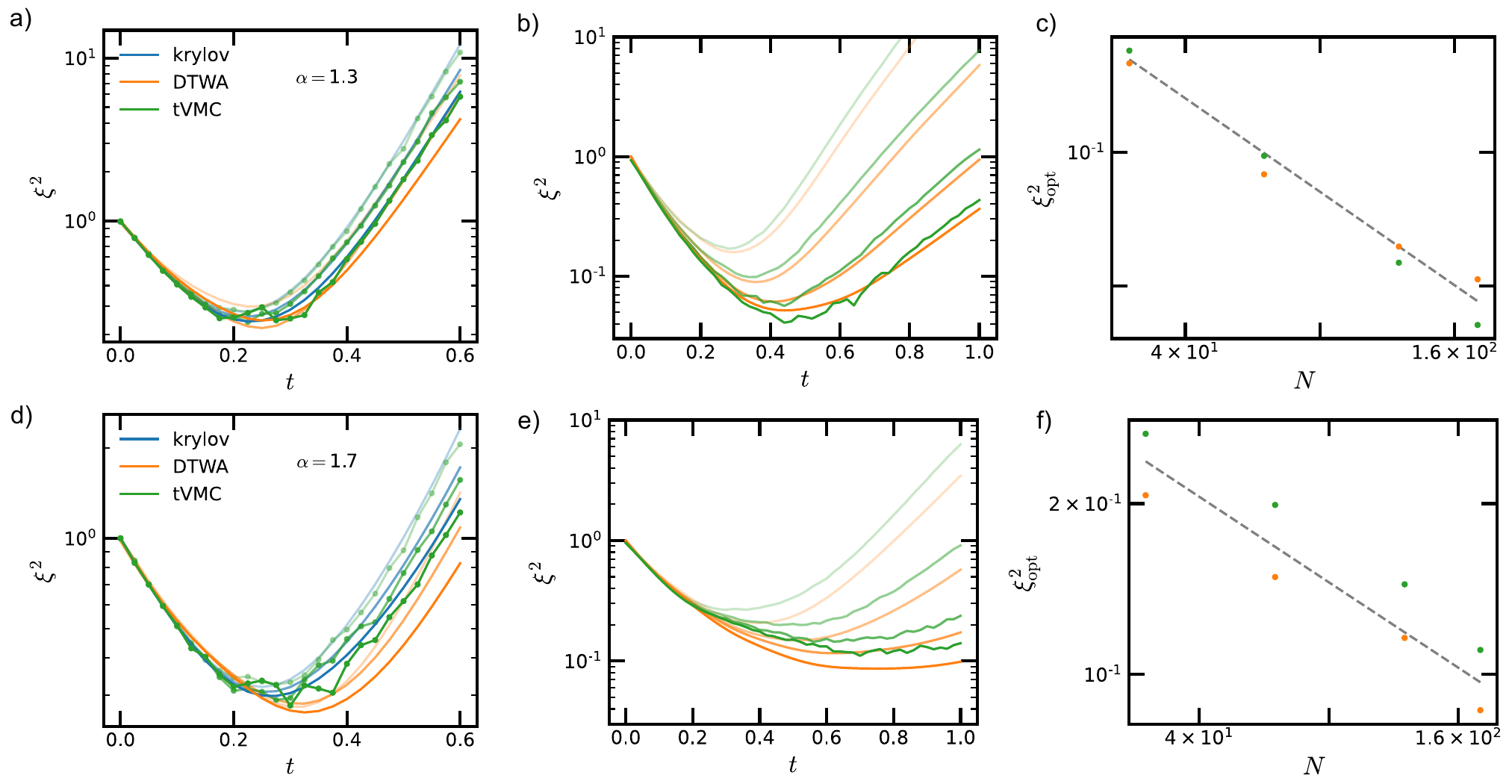}
	\caption{(a) A comparison of squeezing obtained from Krylov-based time evolution, tVMC and DTWA at system sizes $N=14,16,18$ (indicated by increasing opacity) for $\alpha=1.3, J_z=0$ (b) Comparison between tVMC and DTWA at larger system sizes, $N=30,60,120,180$. (c) Comparison of optimal squeezing obtained in (b) as a function of system size. (d-f) Analogous plots for $\alpha=1.7$.}
	\label{fig:bench}
\end{figure}

The resulting comparisons are shown in Fig.~\ref{fig:bench}.
For simplicity, we focus on $d=1$, $\alpha=1.3,1.7$ at $J_z=0$.
Based on our equilibrium phase diagram, we expect $J_z=0$ to be deep in the ordered phase for $\alpha=1.3$ and quite near the critical point for $\alpha=1.7$, so these parameters probe the validity of DTWA in both regimes.
For $\alpha=1.3$ [Fig.~\ref{fig:bench}(a-c)], we find that DTWA agrees well with Krylov across all system sizes studied.
For $\alpha=1.7$, we see that DTWA consistently overestimates the absolute amount of squeezing attained during the evolution (by $\approx 30\%$).
Crucially, however, the \emph{scaling} of squeezing with system size appears to be consistent across methods.
Since the essential aspects of our theory depend only on the scaling of squeezing with system size, it appears that DTWA is a reasonable guide for our analysis even near the critical point. 

Finally, let us also note that in principle, DTWA dynamics cannot accurately capture the short-range spin fluctuations (after local equilibration) deep in the quantum regime, i.e., where the effective temperature of the initial state $T \ll J_\perp$. 
However, this is of little consequence: for $T \ll J_\perp$, finite temperature effects are  important only for very long time-scales at systems sizes beyond our power to investigate. 
Close to the phase transition, where $T$ is of order $J_\perp$, our results suggest that the semiclassical description is more accurate.

\section{Relation to Previous Work}
Generalizing spin-squeezing beyond the one-axis twisting model has been an outstanding challenge in quantum dynamics and several works have conducted seminal investigations and developed valuable insights into the problem.
Here, we we will carefully discuss the relation of our study to these previous works, both to provide an overview of the topic and to delineate our most important contributions.

First, our work was greatly inspired by results of Perlin~et.~al. \cite{perlin}, which showed, using the discrete tructated Wigner approximation (DTWA), that substantial spin-squeezing could be achieved at fixed system size for power-law interacting XXZ systems in $d=2,3$ \footnote{They also look at $d=1$, but only consider integer $\alpha$ and hence only find substantial squeezing in the almost all-to-all case of $\alpha=1$}.
Their results immediately evoke the possibility of a ``squeezing phase diagram," and they even speculate on a possible relationship between squeezing and thermalization to long-range order in their discussion.
However, they conclude that thermalization to a long-range ordered state is an insufficient criterion to fully characterize spin squeezing.
In fact, we argue that thermalization to at long-range ordered state is \emph{the} necessary and sufficient condition that determines whether scalable spin-squeezing will occur.
Our numerical analysis also builds substantially upon that presented in \cite{perlin}.
In particular, we focus not on the quantitative amount of squeezing that can be attained but rather on the qualitative trend of how squeezing scales with system size.
This in turn allows us to define a squeezing phase diagram and hence critically examine the requirements for scalable squeezing.

Second, Comparin et.~al. \cite{comparin} studied spin squeezing in $d=1$ power-law interacting systems using a combination of exact diagonalization and time-dependent variational Monte Carlo (tVMC).
Based on these numerics and analytic approximations, they developed a microscopic explanation for spin-squeezing at small system-sizes based off the large overlap of $\ket{x}$ with the zero-temperature ground-state manifold of the XXZ model, which forms a so-called ``Anderson tower."
In isolation, this manifold behaves as a collective quantum rotor which results in OAT dynamics.
In addition, in agreement with our observations, they find scalable-squeezing ceases abruptly for $\alpha \gtrsim 1.6$ with $J_z=0$.
Based off of their microscopic picture for spin-squeezing, they attribute this to an unfavorable scaling of spin-wave velocity with system size as $\alpha \rightarrow 2$.

While we find their microscopic explanation remarkably insightful and intuitively powerful, we argue there are two related shortcomings of their perspective.
First, thermodynamic considerations suggest their microscopic picture cannot explain scalable spin-squeezing at large system sizes.
Specifically, the finite energy density of $\ket{x}$ implies that it will eventually have no overlap with the zero-temperature Anderson tower, and will be entirely supported on excited states [Fig.~3(f)].
Second, we find that when tuning $J_z$ the transition to scalable spin squeezing varies smoothly and does not remain fixed at $\alpha \approx 1.6$.
However, the scaling of the spin-wave velocity is independent of $J_z$, so it is difficult to explain the $J_z$-dependence of the critical $\alpha$ from this this mechanism.
By contrast, our explanation addresses both of these issues; it accounts for -- and indeed depends crucially on -- the finite temperature of the initial state, and quantitatively predicts the observed scalable-squeezing phase boundary.

Third, a different work by Comparin et.~al. \cite{comparin_scalable_2022} showed that a squeezed state could be prepared by adiabatically reducing a polarizing field in a model with continuous symmetry breaking (CSB) in the ground-state.
Their results complement our work by demonstrating the deep connection between spin-squeezing and CSB in a very different context: the preparation of near zero-temperature equilibrium states.
We remark that their results fit naturally into the framework described in our introduction, and provide another example of how long-range order can serve as the foundation for quantum-enhanced metrology.

Finally, we emphasize that a number of predictions of our work are entirely novel.
Most notably, our claim that squeezing should scale as $N^{^{-\frac{2}{5}}}$ as a consequence of quantum thermalization highlights the fundamental difference between our understanding and previous work.
Relatedly, we observe a transition in scalable squeezing depending on the polarization of the initial state, corroborating our understanding that the squeezing phase transition is determined by the existence of \emph{finite-temperature} order.
Finally, we point out the importance of the conditional ${\rm Var}[Y|Z]$ on squeezing dynamics and define its quantum analogue.

\section{Additional Numerics}

\begin{figure*}
    \centering
    \includegraphics[width=0.95\textwidth]{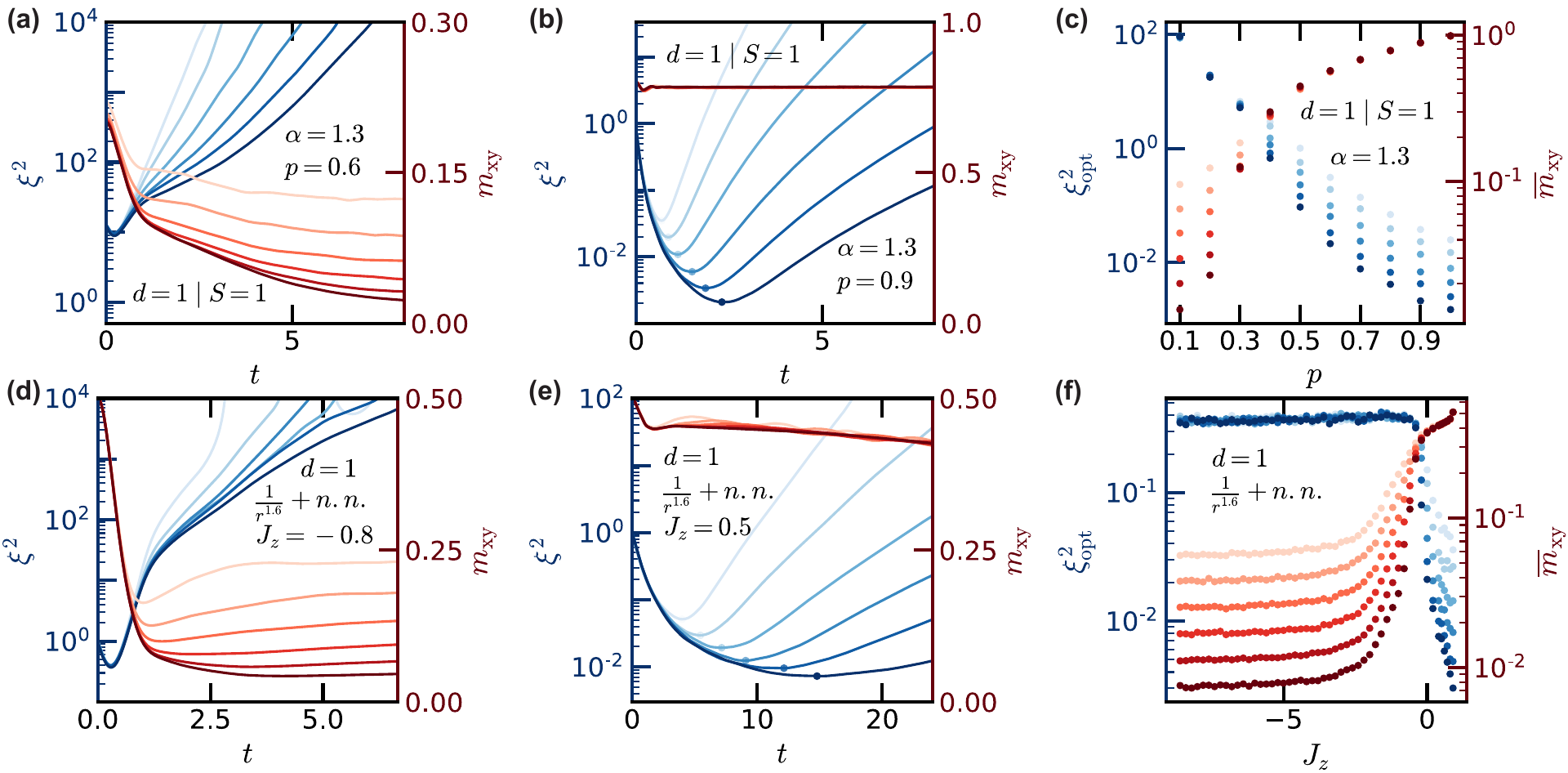}
    \caption{
    (a-c) Squeezing transition as a function of initial polarization $p$ for a $d=1, \alpha=1.3$ spin-$1$ XXZ model, simulated with modified DTWA.
    Following the style of the main text plots,  panels (a) and (b) depict the actual time dynamics in the different phases (i.e. non-squeezing and scalable squeezing), and (c) shows the reciprocity between squeezing scaling and order.
    (d-f) Analogous plots for a $d=1,\alpha=1.6$ spin-$1/2$ XXZ model \textit{with} additional nearest neighbor $XXZ$ interactions. Here, the transition is driven by tuning $J_z$ and polarization is fixed to $1$.
    }
    \label{fig:transition-survey}
\end{figure*}

Here, we provide DTWA numerics supporting our conjecture on two additional $U(1)$ models beyond the spin-$1/2$ Power-law XXZ family, shown in Fig.~\ref{fig:transition-survey}.
Specifically we consider a power-law spin-$1$ XY model (tuning the initial polarization to drive the transition) and a power law spin-$1/2$ XXZ model with extra nearest-neighbor interactions (tuning $J_z$ to drive the transition).
We note that for the spin-$1$ case, we have modified the sampling procedure used in DTWA to match the initial expectation values and covariances of a coherent spin-$1$ state (and have benchmarked the results at small system sizes with exact Krylov methods).
In both cases, we observe a squeezing transition qualitatively identical to that of the power-law XXZ model.
These results demonstrate the squeezing transition is not an accidental feature of power-law spin-$1/2$ XXZ models and thereby emphasize the broad applicability of our theory. 

\bibliography{main-4-z-supp-arxiv.bbl}


\title{Methods and Extended Data for ``Scalable Spin Squeezing from Finite Temperature Easy-plane Magnetism''}

\maketitle

\section{Order and Metrology}

As discussed above, there is a simple connection between SSB order and QFI in pure states, since the variance of the order parameter is nothing but a sum over the connected two-site correlation function.
%
%
Here, we discuss how this relation enables the preparation of metrologically useful states via a Hamiltonian quench: 1) we provide a general argument for why long-range connected correlations develop, which evinces the necessary conditions; 2) we compare our protocol to previous results on the metrological utility of thermal states.

\subsection{General Argument}
We consider the same scheme as described above, where we quench an initial product state with an effective temperature below $T_c$ and assume SSB with order parameter $\mathcal{O}=\sum_i \mathcal{O}_i$.
We further assume $\mathcal{O}$ is purely off-diagonal in the symmetry sectors of the Hamiltonian.
Given this assumption, $\expval{\mathcal{O}}$ must approach $0$ at late times, since all phase coherence between symmetry sectors will be lost.
On the other hand, $\mathcal{O}^2$ has some diagonal elements that act only within symmetry sectors.
Since we are assuming that the initial state has effective temperature below $T_c$, the expectation value of these diagonal elements should be $O(N^2)$ \cite{deutsch_eigenstate_2018,dalessio_quantum_2016}.
This leads to the conclusion $\langle \mathcal{O}^2 \rangle - \langle \mathcal{O} \rangle^2 \sim O(N^2)$, i.e. the late time state exhibits a large QFI with respect to $\mathcal{O}$.
Note that if $\mathcal{O}$ is \emph{diagonal} in the symmetry sectors, then $\langle \mathcal{O} \rangle$ can be large at all times and this expression can be small (i.e. $O(N)$).

For the above argument to be physically relevant, it must further be shown that that the ``late time" at which $\expval{\mathcal{O}}$ approaches $0$ does not diverge too rapidly in system size.
In the most general case, it is difficult to analyze this time scale based on the symmetry of $H$ alone.
However, assuming the finite-temperature SSB derives from zero-temperature SSB of the same kind, e.g. long-range interactions stabilizing CSB in $1D$ or $2D$, one can plausibly estimate this time scale from the ground-state manifold itself. 
Specifically, models with discrete symmetry have exponentially small finite-size gaps between ground-states, leading to an exponentially long dephasing time between symmetry sectors.
On the other hand, models with continuous symmetry breaking, such $U(1)$, should exhibit $\sim O(1/N)$ gaps between symmetry sectors, so the dephasing time scale between sectors is only linear in system size and QFI develops relatively quickly.

To make the limitations of discrete symmetry palpable, let us consider the power-law generalization of the two-axis counter twisting model, which has only discrete symmetry,
\begin{equation}
    H_{\alpha} = \sum_{i,j} \frac{1}{r_{ij}^\alpha} (\sigma^y_i \sigma^y_j - \sigma^z_i \sigma^z_j).
\end{equation}
While the all-to-all interacting version of this model yields scalable spin squeezing for the initial state $\ket{x}$, the above version does not. In light of the above discussion, the $\ket{x}$ state must either exceed the critical temperature, or the dynamics between the symmetry sectors will be exponentially slow. It turns out that $\ket{x}$ is above the critical temperature. To see this, note that $\ket{x}$ has $0$ energy density and the spectrum of $H_\alpha$ is symmetric about $E=0$; thus $\ket{x}$ is at infinite (effective) temperature. The absence of scalable squeezing in this model is substantiated by DTWA numerics (Extended Data Fig.2).

We note that the above arguments require few body operators to thermalize quickly and thus do not apply to integrable models, to models that have anomalously slow thermalization for certain initial states (e.g., quantum many-body scars), or to models with super-extensive energy ($\alpha \leq d$) \cite{defenu_long-range_2021}.
Assuming conventional local thermalization, the key conditions for generating large QFI from a quench are therefore 1) a Hamiltonian with continuous symmetry breaking \emph{and} an order parameter off-diagonal in the symmetry sectors (See the Supplementary Information for an example to illustrate the importance of this requirement~\cite{supp}) and 2) the existence of a low-temperature symmetry breaking initial state that is easy to prepare, e.g., a product state. 


\subsection{QFI in thermal states}

Strictly speaking, the equivalence between QFI and order only holds for pure states, and can only be utilized via coherent quench dynamics or adiabatic preparation of a symmetry breaking state \cite{comparin_scalable_2022}.
As stated above, thermal density matrices (even at equivalent energy density) have significantly reduced metrological utility and do not generally enable quantum enhanced sensing. 
To gain some intuition about this distinction, consider that for large systems the low-temperature thermal density matrix will be well approximated by an incoherent mixture of all low-energy symmetry-breaking states. 
This mixture of essentially classical states clearly cannot provide a quantum enhancement to the sensitivity~(Extended Data Fig.3).
This intuition is made rigorous in recent studies of the QFI of thermal states, which show that only the finite size gap above the ground state suppresses this inimical incoherent mixing \cite{hauke_measuring_2016,gabbrielli_multipartite_2018}.

Consequently, an experimental realization of our proposal must contend with limits imposed by decoherence.
For sufficiently large systems, any amount of decoherence will ultimately cutoff beyond-SQL metrological gain.
In practical terms however, this threshold can be quite high.
Indeed, several experimental platforms have coherence times of $\gtrsim 10^2/J$, where $J$ is the local energy scale \cite{PRXQuantum.2.017003}.
Since optimal-squeezing arises at sub-extensive times, this implies that at least one or two orders-of-magnitude  of beyond-SQL scaling may be observed before decoherence is severely limiting.
We also note that certain types of incoherent dynamics, e.g., starting with a partially-polarized mixed state, do not necessarily undermine the feasibility of our quench protocol (see section \ref{sec:mixed-init} below)

Finally, see Extended Data Table 1 for a summary of how various forms of the connected correlation function result in different degrees of enhancement over the SQL.

\section{Extracting the squeezing phase diagram}

According to our analysis, the non- squeezing and scalable squeezing phases should be distinguished by a sudden change of the exponent $\nu$ from 0 to 2/5, where $\nu$ is defined by the large $N$ behavior of the optimum squeezing, $\xi_{\rm opt}^2 \sim N^{-\nu}$.
However, two types of finite-size effects make it difficult to observe this directly in numerics.

The first challenge is that the system always undergoes some \emph{non-scalable} squeezing at short times, as shown in Extended Data Fig.4,Fig.5,Fig.6 (see also ~\cite{roscilde_bilinear}). 
In a regime of modest easy-axis coupling and small system sizes, this early time minimum can actually be quantitatively better than the late time minimum that \emph{scales} with system size.
To properly evaluate the squeezing transition, we need to isolate the late-time, scalable squeezing.
To this end, we introduce a more nuanced definition of optimal squeezing, $\xi_{\rm opt}^2$, as the smallest value of $\xi^2(t)$ at a local minima in the derivative $\frac{d}{dt} \xi^2(t)$ after the local thermalization time.
In the squeezing phase, this definition coincides with the true minimum of $\xi^2(t)$. 
In the non-squeezing phase, no features of $\xi^2(t)$ scale with system size so $\nu=0$ regardless of the criteria for selecting a representative squeezing value. 
Thus, our definition only impacts the observed scaling near the critical point, where it serves to mitigate the influence of the aforementioned early-time squeezing on the analysis.

The second significant finite-size effect is that the linear growth of $\textrm{Var}[Y|Z]$ may not affect the squeezing scaling until one examines very large system sizes (which achieve optimal squeezing at very late times).
In more detail, the slope of $\textrm{Var}[Y|Z]$ (in units of its early time value) decreases with effective temperature (in units of the local energy scale).
For $J_z \gtrsim 0$, the effective temperature of $\ket{x}$ is so low that one may not see $N^{-2/5}$ scaling until system sizes of millions or tens of millions of spins.
Hence, to observe $\nu=2/5$ in numerics we resort to tuning the temperature \emph{without} changing the local energy scale, by altering the polarization of the initial state [Fig.~3(b,c)]. 
By contrast, if one increases the temperature by tuning $J_z$, the local energy scale will simultaneously increase, maintaining the system-size threshold for observing $N^{-2/5}$ scaling at numerically inaccessible system sizes.

We emphasize that, assuming the validity of our semiclassical model, the squeezing \emph{will} scale as $N^{-2/5}$ for sufficiently large systems, since we \emph{do} see linear growth of $\textrm{Var}[Y|Z]$ and its quantum analog [Fig.~3(e)].
It is just not apparent at the system sizes we are able to simulate.
Therefore, it is not appropriate to look for $\nu=2/5$ exactly as a signature of the phase transition.
Indeed, in our DTWA numerics to study the transition as a function of $J_z$ in $d=1,2$ from $N=[120...10379], L=[11...102]$, starting from the fully polarized $\ket{x}$ state, we essentially always see $\nu=2/3$ in the squeezing phase.

The above effects make it difficult to propose a definitive scaling function for $\nu$ that one could use perform standard finite-size scaling. 
In lieu of this, we analyze the scaling of the of optimal squeezing \emph{time}, where $t_{\rm opt} \sim N^\mu$, in addition to the scaling of the optimal squeezing \emph{value}, to obtain a more robust estimate of the squeezing critical point (Extended Data Fig.7).
Specifically, we estimate as the $J_z$ value at which $\mu$ and $\nu$ become significantly greater than $0$.
In more detail, we subdivide the squeezing into $25$ ``bootstrap" subsets.
For each subset, we determine the optimal squeezing time and and value, as functions of $(\alpha, J_z, N)$, in the manner described above.
From this, we obtain, for each $\alpha$, a sample of $\nu(J_z), \mu(J_z)$.
These are fit by independent logistic curves, and the left elbow of each logistic is used as an estimate of $J_c$.
The intervals containing all such $J_c$ are denoted by purple errorbars in Fig.~1(b,c), and their midpoints are marked as estimates of $J_c$ for the squeezing transition.
As emphasized above, the critical interval closely follows the thermal critical point, defined as the $J_z$ at which the effective temperature of $\ket{x}$ exceeds $T_c$ for CSB order.


In $d=3$, $\nu$ jumps discontinuously from $\approx 0$ at $J_z=-1.0$ to $\approx 0.45$ at $J_z=-0.9$, so the squeezing critical point clearly lies in this interval (according to our conjecture, it should be exactly $J_z=-1.0$).
%

\section{Hydrodynamic model for squeezing dynamics}
As discussed in the main text, spin-squeezing dynamics are well described by a semi-classical picture that approximates the quasi-probability distribution as a true probability distribution evolving on the global Bloch sphere. 
%
%
%
%
Here, based on a hydrodynamic description of spontaneous symmetry breaking, we analytically derive the equations of motion for the probability distribution (which we further use to calculate the optimal squeezing in the next section). 
%
Our main assumption is that after a relatively short time (of order $1/J$), the system achieves local thermal equilibrium and the subsequent dynamics of coarse-grained variables can be described by hydrodynamic equations.

The hydrodynamic description focuses on two  coarse-grained variables that evolve only slowly in time at long wavelengths,  the $z$-component of the magnetization density, denoted $m(\mathbf{r}) $ (the $U(1)$ charge), and an  angle $\phi(\mathbf{r})$ describing the orientation  of the magnetization  in the x-y plane.  The magnitude  of the in-plane component, $m_{\mathrm{xy}}$, is assumed to have relaxed rapidly to a value determined by the temperature $T$,  which we take to be uniform in space. The phase $\phi$ will vary slowly at long wavelengths because the restoring force vanishes for long wavelength fluctuations, and $m$ varies slowly because it is a conserved quantity.
A hydrodynamic description should apply for sufficiently long wavelengths at any non-zero temperature, regardless of whether the underlying microscopic system obeys classical or quantum dynamics.

 Integrating the partition function over short range fluctuations in $m$ and $\phi$, and over all other variables in the problem, we may define a free energy functional, which we assume to take the form
 \begin{equation}
   F [m, \phi] = \int d^d r \,  \left( \chi   m(\mathbf{r}) ^2 + u_4 m(\mathbf{r})^4   + ...   \right)  + \int d^d r_1 d^d r_2   \tilde{J} (\mathbf{r}_1 - \mathbf{r}_2 )   \left\{ \cos  [\phi(\mathbf{r}_1)-\phi(\mathbf{r}_2)] \right\}  ,  
 \end{equation}
 where $\tilde{J} (\mathbf{r}) \sim ( J_\perp /2)   m_{\mathrm{xy}}^2  r^{- \alpha}  $ for large separations $r$. 
 The hydrodynamic equations of motion take the form
\begin{equation}
\begin{split}
    \frac{\partial \phi}{\partial t} &= g\frac{\delta F}{\delta m}-\Gamma \frac{\delta F}{\delta \phi} + \eta_\phi(t,\vec{r}),\\
    \frac {\partial m}{\partial t} &=  - g  \frac {\delta F}{\delta \phi} + \nabla^2 \left[ \Lambda \frac {\partial F} {\partial m} + \eta_m(t,\mathbf{r}) \right], \\
    \langle \eta_{\phi}(t,\vec{r}) \rangle &= \langle \eta_{\phi}(t,\vec{r}) \rangle = 0, \\
    \langle \eta_\phi(t,\vec{r}) \eta_\phi(t',\vec{r}') \rangle &= 2\Gamma T \delta(t-t')\delta(\vec{r}-\vec{r}'), \\
    \langle \eta_m(t,\vec{r}) \eta_m(t',\vec{r}') \rangle &= 2\Lambda T \delta(t-t')\delta(\vec{r}-\vec{r}'), 
\end{split}
\label{eq:hydro}
\end{equation}
where $\eta_\phi$ and $\eta_m$ originate from the thermal fluctuations.
%
Here we define the global quantity $\Phi|Z$, where 
\begin{equation}
    \Phi=\frac{\int d^d\vec{r} \phi(\vec{r})}{V}
\end{equation}
is the phase averaged over the entire system, given fixed total magnetization along z-axis, 
\begin{equation}
    \int m(\vec{r})~d^d r=Z. 
\end{equation}
\noindent The coefficient $g$ in the hydrodynamic equations is determined by the commutator of $m$ and $\phi$. In the present problem, $g=1$.   
Our hydrodynamic equations are a generalization to the case of long-range spin interactions of the hydrodynamic equations that are commonly used for an XY magnet with short-range interactions in two or more dimensions. \cite{halperin_hydrodynamic_1969}\cite{hohenberg_theory_1977}. It is expected that these equations should be valid at any temperature in the broken symmetry phase at sufficiently long wave lengths. The hydrodynamic equations are also believed to be valid in the low-temperature phase of the two-dimensional system with short range forces, where there is no true broken symmetry but only quasi-long-range order of the x-y magnetization. We do not have a rigorous derivation of these equations starting from the microscopic Hamiltonian, but one can at least perform some consistency checks.  For example, one can confirm that non-linear coupling to thermally excited long-wavelength spin fluctuations does not lead to a divergence of the coefficient $\Gamma$. We note that the equations of motion conserve the value of $M$, and they are consistent with a time-independent thermal distribution of the form
 $ P[m, \phi]  = \mathcal{Z}^{-1}  e ^{- F[m,\phi]/ T }$ .

For small fluctuations about the equilibrium state, where $m=0$ and $\phi$ is independent of space, one can solve the dynamics  of Eq.~\eqref{eq:hydro} by performing a Fourier transform. This leads to  the following equations of motion in momentum space
\be
\label{dphikdt}
\frac {d \phi _{\bfk}} {dt} = 2 g \chi  m_\bfk - \Gamma K_\bfk \phi_\bfk + \eta_{\phi, \bfk}(t) 
\ee
\be 
\frac {d m_\bfk} {dt} = - g K_\bfk \phi_\bfk -  k^2 [ 2  \Lambda \chi m_\bfk + \eta_{m, \bfk} (t) ] ,
\ee
where $K_\bfk$ is related to $J_\bfk$, the Fourier transform of $\hat{J}$, by
\be
K_\bfk = 2 ( J_0 -J_\bfk).  
\ee
The noise terms satisfy
\be
\ex {\tilde { \eta}_{\phi, \bfk} (t)    \tilde{ \eta}_{\phi, \bfk}^*  (t')   }  = \frac {2 \Gamma T }{N} \delta ( t-t')  ,  \,\,\,\,\, 
\ex {\tilde { \eta}_{m, \bfk} (t)    \tilde{ \eta }_{m, \bfk}^*  (t')   }  = \frac {2 \Lambda T }{N} \delta ( t-t')  .
\ee

For small wave vectors, one finds  $K_\bfk \sim \tilde{K} k ^{\alpha - d} $ , where the coefficient $\tilde{K}$ is proportional to  $J_\perp m_{\mathrm{xy}}^2$.   Then, for $k \neq 0$, the equations of motion lead to  propagating spin waves,  with frequency $\omega_\bfk \sim ( 2 g ^2   \chi  K_\bfk)^{1/2}$  and a damping rate proportional to  $ \Gamma K_\bfk + 2 \Lambda \chi k^2$.

Crucially, the $k=0$ mode is different from the other modes. The first term in Eq.~\eqref{dphikdt} just gives a constant precession rate, $2Z\chi/N$ due to fixed total magnetization, and the second term also vanishes  for $\alpha >  d$.  The equation of motion for $\Phi$ then reads
\be
\frac {d \Phi}{dt} = \frac{2Z\chi}{N}+\tilde{\eta} _\phi (t). 
\ee
Thus the mean value of the precession angle grows linearly,
\be
\Phi(t) |Z = \frac{2Z\chi t}{N}. 
\label{eq:phi_mean_dynamics}
\ee
Whereas the second term, which drives a random walk, leads to linear growth of the variance in a fixed $Z$ sector: 
\begin{equation}
{\rm{Var}} [\Phi (t)|Z] = {\rm{Var}} [\Phi (0) ] + \frac {2 \Gamma T}{N} t. 
\label{eq:phi_var_dynamics}
\end{equation}
We note that the spin operators $X,Y,Z$ and $\Phi$ are related by:
\begin{equation}
\begin{split}
X|Z &= N m_{\mathrm{xy}} \mathrm{cos}(\Phi|Z),\\
Y|Z &= N m_{\mathrm{xy}} \mathrm{sin}(\Phi|Z).
\end{split}
\label{eq:phi_xy}
\end{equation}
And therefore 
\begin{equation}
{\rm Var}[Y|Z] \approx m_{\mathrm{xy}}^2 N^2 \mathrm{Var}[\Phi|Z],
\end{equation}
for small mean precession angle, which holds for $t \ll \Gamma T N$.

Two remarks are in order. 
First, the linear growth rate vanishes when temperature $T\rightarrow 0$, suggesting that such effect is very weak for the quench dynamics from a low-temperature initial state. 
This explains why our proposed scaling behavior of squeezing is hard to observe at low temperature. 
We note that the linear growth rate will probably vanish \emph{faster} then $T$ at low temperature, since the noise strength $\Gamma$ can also depend on temperature and vanish itself when $T\rightarrow 0$. 
Second, the behavior of the non-zero momentum modes ($k\neq 0$) is qualitatively different from the zero-momentum mode: due to the non-zero restoring force, $\mathrm{Var}[\phi_k(t)]$ cannot increase to infinity and will instead saturate to a temperature-dependent equilibrium value. 
This effect manifests as the evolution towards local equilibration in the quench dynamics. 

We remark that the analysis described above is restricted to the case where the z-component of the total magnetization is confined to a small interval about $Z=0$. At finite value of $Z/N$, the hydrodynamic description must take into account fluctuations in energy density, which gives rise to an additional slow mode due to energy conservation. Energy fluctuations couple linearly to the spin modes for $Z/N \neq 0$, and this coupling can have significant effects. Because the initial state is a superposition of states with different energies, and these energies persist to infinite times, we expect that in the large $N$ limit, the quantity Var$[\Phi(t)]$ should grow as $t^2  / N$ for  large $t$.  This reflects the fact that systems with slightly different total energies will have slightly different values of the parameter $\chi$.  For a state where $Z/N$ is of order $N^{-1/2}$, however, the quantity Var$[\Phi(t)]$ should grow as $t^2  / N^2$ for  large $t$, which is too small to affect the squeezing behavior.

\section{Derivation of the squeezing parameter via semi-classical approach}
%
Following the previous section, we now present an analytical calculation of the optimal squeezing based on the equations of motion for the probability distribution. 
In particular, we derive Eq.~(2) of the main text. 

The expression of the spin squeezing parameter $\xi^2$ consists of the mean spin length $\langle X\rangle$ and the minimum variance $\min_{\hat{n}\bot \hat{x}}{\mathrm{Var}[\hat{n}\cdot \vec{S}]}$ in the y-z plane. 
The latter is of course the smallest eigenvalue of $Y$, $Z$ covariance matrix,
\begin{equation}
\begin{pmatrix}
\langle Z^2\rangle & \langle ZY\rangle\\
\langle YZ\rangle & \langle Y^2\rangle
\end{pmatrix}. 
\label{eq:covmatrix}
\end{equation}
Therefore, to calculate the squeezing we only need to evaluate the observables $\langle X\rangle$, $\langle Z^2\rangle$, $\langle Y^2\rangle$ and $\langle ZY\rangle=\langle YZ\rangle$.


Since $Z$ is a conserved quantity, we can consider the evolution of each $Z$-slice of the  probability distribution separately.
The population within each slice is conserved and determined by the initial binomial distribution of $Z$, which can be well approximated by a Gaussian distribution in the thermodynamic limit. That is,
\begin{equation}
P(Z)=\sqrt{\frac{2}{\pi N}}\mathrm{e}^{-\frac{2Z^2}{N}}. 
\end{equation}
As per Eq.~\ref{eq:phi_mean_dynamics} and Eq.~\ref{eq:phi_xy}, the dynamics in each $Z$-slice are described by a rotation with a angular velocity of $\frac{2Z\chi}{N}$ given as the follows:
\begin{equation}
\begin{split}
    X|Z &= N m_{\mathrm{xy}} \mathrm{cos}\left(\frac{2Z \chi t}{N}\right),\\
    Y|Z &= N m_{\mathrm{xy}} \mathrm{sin}\left(\frac{2Z \chi t}{N}\right). 
\end{split}
\end{equation}
Therefore, all the terms in $\xi^2$ can be evaluated as 
\begin{equation}
\begin{split}
    \langle X\rangle &= \int N m_{\mathrm{xy}} \mathrm{cos}\left(\frac{2 Z \chi t}{N}\right) P(Z)dZ = N m_{\mathrm{xy}} \mathrm{e}^{-\frac{(\chi t)^2}{2N}}\\
    \langle Z^2\rangle &= \int Z^2 P(Z)dZ = \frac{N}{4}\\
    \langle ZY\rangle &= \langle YZ\rangle = \int Z N m_{\mathrm{xy}} \mathrm{sin}\left(\frac{2 Z \chi t}{N}\right) P(Z)dZ = \frac{1}{2} N m_{\mathrm{xy}} \chi t~\mathrm{e}^{-\frac{(\chi t)^2}{2N}},\\
    \langle Y^2\rangle &= \mathrm{Var}[Y|Z]+\int N^2 m_{\mathrm{xy}}^2 \mathrm{sin}^2\left(\frac{2 Z \chi t}{N}\right) P(Z)dZ \\
    &= \mathrm{Var}[Y|Z]+\frac{1}{2}N^2 m_{\mathrm{xy}}^2[1-\mathrm{e}^{-\frac{2(\chi t)^2}{N}}]. 
\end{split}
\end{equation}
Plugging the above last three lines into Eq.~\ref{eq:covmatrix}, we obtain the minimum variance in the y-z plane as
\begin{equation}
\begin{split}
    \min_{\hat{n}\bot \hat{x}}{\mathrm{Var}[\hat{n}\cdot \vec{S}]} &= \frac{1}{2}\left\{\mathrm{Var}[Y|Z]+\frac{N}{4}+\frac{1}{2}N^2 m_{\mathrm{xy}}^2[1-\mathrm{e}^{-\frac{2(\chi t)^2}{N}}]\right\}\\
    &-\frac{1}{2}\left[\left\{\mathrm{Var}[Y|Z]-\frac{N}{4}+\frac{1}{2}N^2 m_{\mathrm{xy}}^2[1-\mathrm{e}^{-\frac{2(\chi t)^2}{N}}]\right\} ^2+N^2m_{\mathrm{xy}}^2\chi^2 t^2\mathrm{e}^{-\frac{(\chi t)^2}{N}}\right]^{1/2}. 
\end{split}
\end{equation}

Scalable squeezing occurs at later and later times as $N$ increases, but occurs \emph{before} the quantum fisher information reaches maximum at $\sim \sqrt{N}$ \cite{roscilde_cat_states}. Hence we consider the following limit
\begin{equation}
    \chi t \rightarrow \infty,\qquad\quad    \frac{\chi t}{\sqrt{N}} \rightarrow 0, 
\end{equation}
and expand the minimum variance in series
\begin{equation}
    \min_{\hat{n}\bot \hat{x}}{\mathrm{Var}[\hat{n}\cdot \vec{S}]} = \frac{N}{4}\left\{\frac{{\rm Var}[Y|Z]/N}{m_\textrm{xy}^2 (\chi t)^2}+\frac{(\chi t)^4}{6N^2}+\mathcal{O}\left[\frac{1}{(\chi t)^2}\right] +\mathcal{O}\left[(\frac{\chi t}{\sqrt{N}})^4\right] \right\}. 
\end{equation}
We note the mean spin length is simply constant in this limit:
\begin{equation}
    \langle X\rangle ^2 = N^2 m^2_\textrm{xy}\left\{1+\mathcal{O}\left[(\frac{\chi t}{\sqrt{N}})^2\right]\right\}
\end{equation}
Combining the above, we obtain Eq.~(2) in the main text. 

Let us finally remark that the time dependence of the conditional variance ${\rm Var}[Y|Z]$ fully determines the scaling behavior of $\xi_{\mathrm{opt}}$ and $t_{\mathrm{opt}}$ as a function of $N$. 
Assuming ${\rm Var}[Y|Z]\propto N(\chi t)^\gamma$ (with $0\leq \gamma<2$), by minimizing Eq.~(2) we expect
\begin{equation}
    \xi_{\rm{opt}} \propto N^{-2+\frac{8}{6-\gamma}},\qquad    t_{\rm{opt}} \propto N^{\frac{2}{6-\gamma}}. 
\end{equation}
In particular, for OAT model, ${\rm Var}[Y|Z]$ is constant in time, i.e. $\gamma = 0$, leading to the scaling behavior $\xi_{\rm{opt}} \propto N^{-\frac{2}{3}}$ and $    t_{\rm{opt}} \propto N^{\frac{1}{3}}$. 
In contrast, for the ordered phase in a thermalizing model, as shown in Eq.~\ref{eq:phi_var_dynamics} and Eq.~\ref{eq:phi_xy}, we expect ${\rm Var}[Y|Z] = 2 N m_{\rm{xy}}^2 \Gamma T\cdot t\propto t$, i.e. $\gamma = 1$. 
Correspondingly, we predict a different scaling behavior $\xi_{\rm{opt}} \propto N^{-\frac{2}{5}}$ and $    t_{\rm{opt}} \propto N^{\frac{2}{5}}$. 

\section{Scalable Squeezing with Mixed Initial States} \label{sec:mixed-init}
Here we elaborate on the order-of-averaging distinction between calculating squeezing for partially-polarized random pure states vs.~a partially polarized mixed state.
In the first case, we fix a product state $\ket{\psi(p)}$, where each spin is independently chosen to point in $\pm \hat{x}$ with probability $(1\pm p)/2$.
To determine the typical value of the squeezing given $p$, we average over the squeezing parameter for each $\ket{\psi(p)}$,
\begin{equation}
\xi^2_{p} =  \mathbb{E}_{\psi(p)} \biggl[ \frac{\min_{\hat{n}\bot \hat{x}}{\mathrm{Var}_{\ket{\psi(p)}}[\hat{n}\cdot \vec{S}]}}{\bra{\psi(p)}X\ket{\psi(p)}^2} \biggr]
\end{equation} 

For a mixed initial state of the form $\rho = \left[ (1+p)/2 \ket{\rightarrow}\bra{\rightarrow} + (1-p)/2\ket{\leftarrow}\bra{\leftarrow}\right ]^{\otimes N}$, the trace over $\rho$ to compute expectation values again becomes an average over $\ket{\psi(p)}$.
This leads to a squeezing parameter of 
\begin{equation}
\xi^2_{{\rm mix}} = \frac{\min_{\hat{n}\bot \hat{x}}{\mathrm{Var}_{\rho}}[\hat{n}\cdot \vec{S}]}{[\mathrm{Tr}(\rho X)]^2} = \frac{\mathbb{E}_{\psi(p)}[\min_{\hat{n}\bot \hat{x}}{\mathrm{Var}_{\ket{\psi(p)}}[\hat{n}\cdot \vec{S}]}]}{\mathbb{E}_{\psi(p)}[\bra{\psi(p)}X\ket{\psi(p)}]^2}
\end{equation} 

Note that squeezing is now calculated \emph{after} the averaging.
Due to the non linearity of the squeezing parameter, $\xi^2_p$ and $\xi^2_{\rm mix}$ could, in principle, be very different (and even have different scaling properties).
However, these quantities agree extremely well for even modestly sized systems $N \gtrsim 10^2$.
This is because, for large systems, mean spin direction becomes essentially a constant -- it is $\sim pN + \mathcal{O}(\sqrt{N})$, and decays a negligible amount before the optimal squeezing time.  

\bibliography{main-4-methods-arxiv.bbl}



\begin{table}[h!]
\centering
\begin{tabular}{cccc}
    \hline
    \makecell{\textbf{Asymptotic} \\ \textbf{Correlation} \textbf{Function}} & \makecell{\textbf{QFI}\\ \textbf{Scaling}} & \makecell{\textbf{Sensitivity}\\ \textbf{Scaling}} \\
    \hline
    \hline 
    $\sim e^{-\abs{x-y}/\xi}$ & $N$ & $\frac{1}{\sqrt{N}}$ \\
    \hline
    $\sim \abs{x-y}^{-p}$ with $p\geq d$ & $N$ & $\frac{1}{\sqrt{N}}$ \\
    \hline
    $\sim \abs{x-y}^{-p}$ with $p < d$ & $N^{2-p/d}$ & $\frac{1}{N^{1-p/2d}}$ \\
    \hline
    $\sim \textrm{const.}$ & $N^2$ & $\frac{1}{N}$ \\ 
    \hline
\end{tabular}
\caption{Relation between order, QFI and sensitivity in the case of short range, quasi long-range order, and true long-range order. }
\label{tab:order-QFI}
\end{table}

\begin{figure*}[h!]
    \centering
    \includegraphics[width=0.65\textwidth]{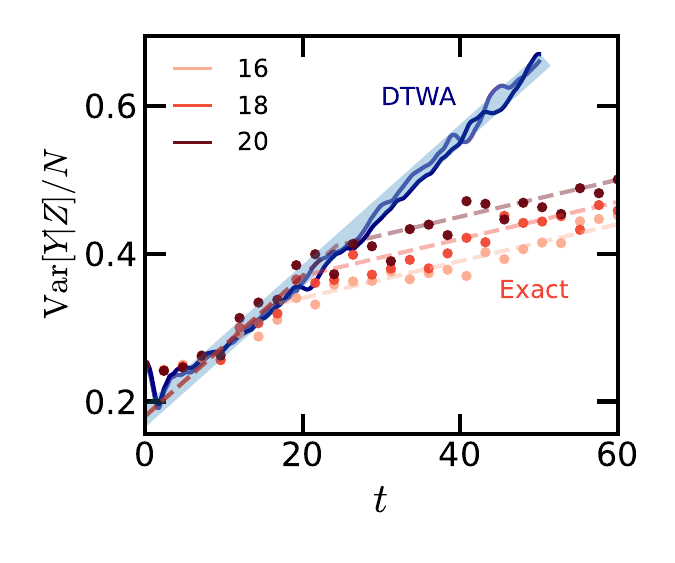}
    \caption{Demonstration of long-time linear growth of $\mathrm{Var}[Y|Z]$. Large-scale ($N=1600,2400$) DTWA results up to times $t \sim 50$ (blue) show consistent linear growth as predicted by hydrodynamics. Exact simulations of small systems using Krylov subspace methods (red) give almost identical results at intermediate times. At later times, the exact results differ from the hydrodynamic expectation due to strong finite size effects arising from coherences between states of adjacent magnetization sectors.
    }
    \label{fig:linear-growth}
\end{figure*}

\begin{figure}
	\centering
 \includegraphics[width=0.8\textwidth]{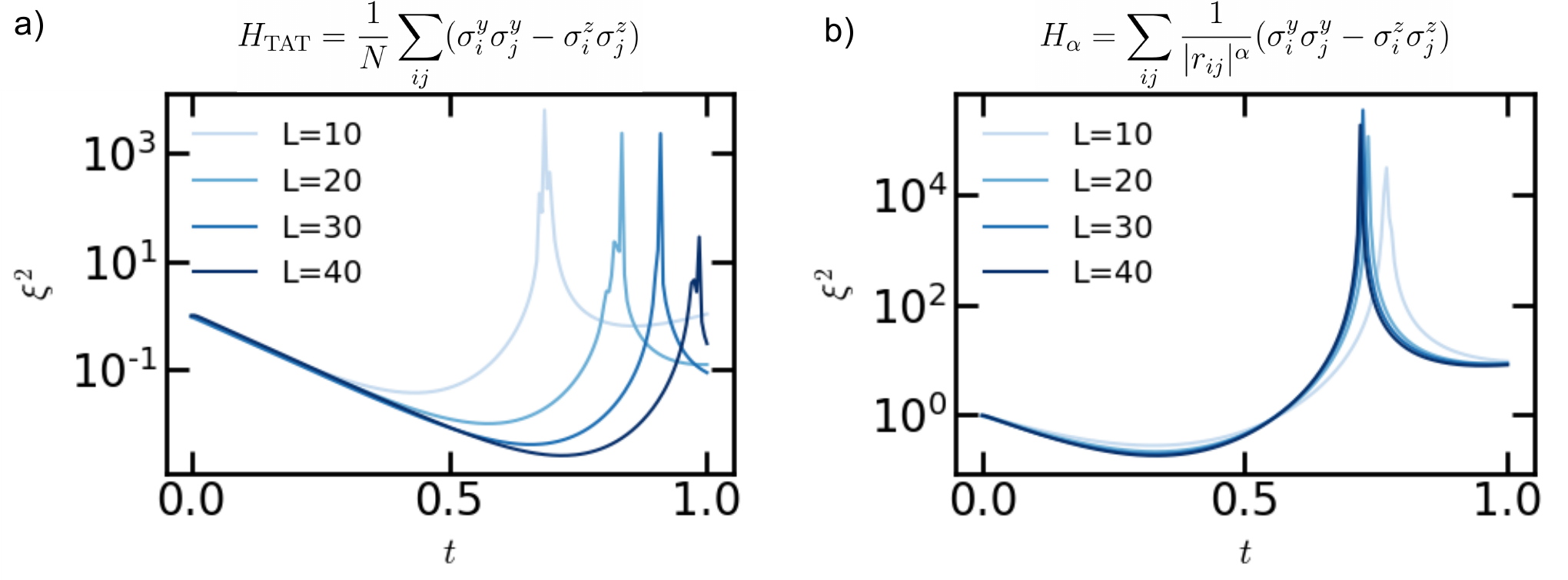}
	\caption{A comparison of TAT dynamics in all-to-all (left) vs power-law interacting (right) systems. The power-law generalization of the TAT model \emph{does not} enable scalable spin squeezing. Here, the initial state $\ket{x}$ is simply at too high a temperature for $\mathbb{Z}_2$ order, preventing any quantum enhancement of the QFI. However, we emphasize, as discussed in the Methods, lowering the temperature of the initial state \emph{would not} restore spin squeezing.}
	\label{fig:TAT}
\end{figure}



\begin{figure}
    \centering
    \includegraphics[width=0.95\textwidth]{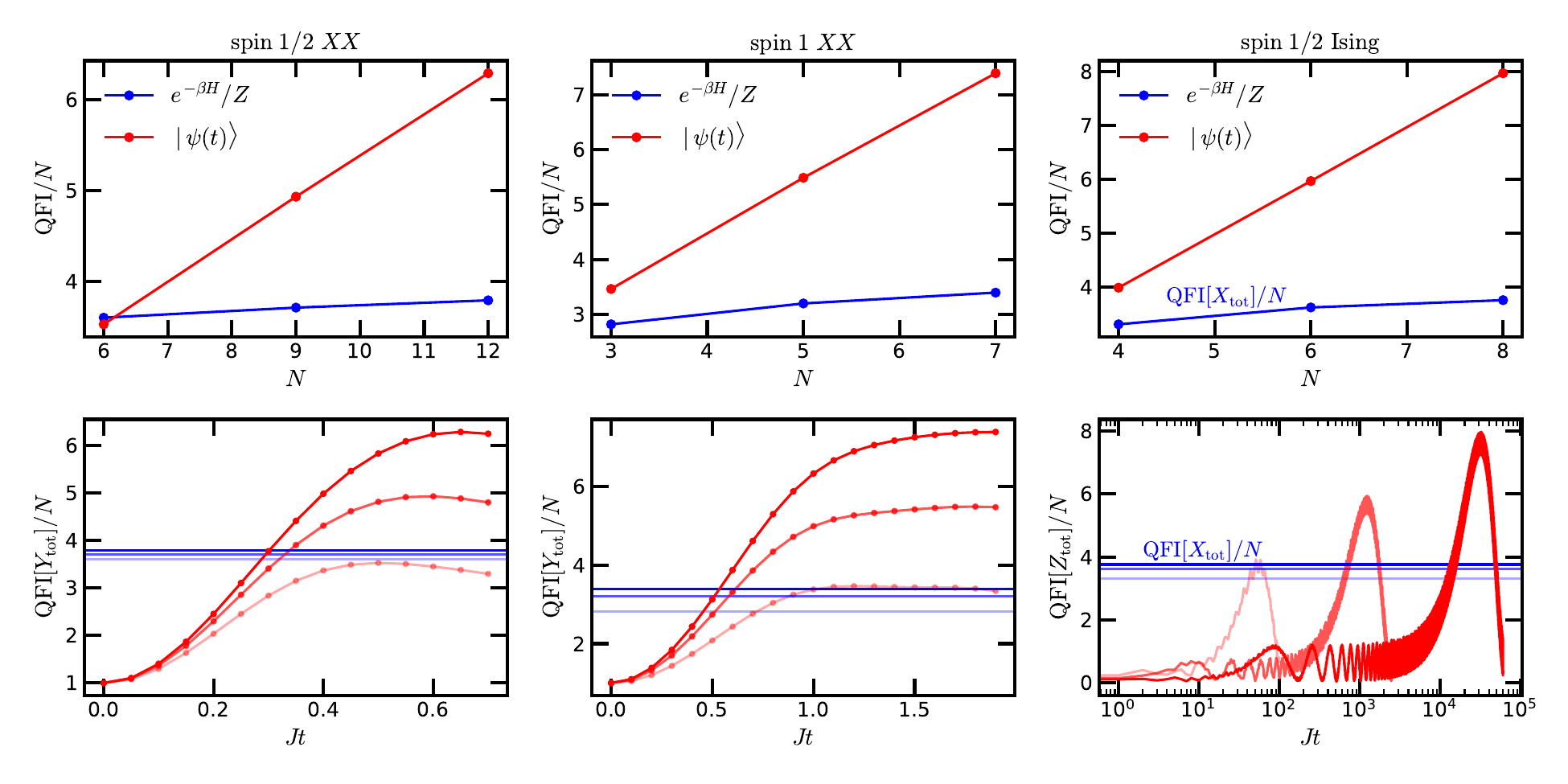}
    \caption{A comparison between the QFI density of the thermal density matrix (blue) and of a product state evolving after a quench (at the same energy density, red) for three models (from left to right): a $d=1$ spin-$\frac 1 2$ a long-range XX model ($\alpha=1.7, J_z=0, J_\bot=-1.0, N=[6,9,12]$), a spin-$1$ version of the same model ($N=[3,5,7]$) and a $d=1$ spin-$\frac 1 2$ a long-range Ising model ($\alpha=1.3, J_z=-1.0, g=0.7, N=[4,6,8]$). The top row shows the maximum QFI of the product state under quench dynamics as function of system size; the bottom row shows the corresponding dynamics, with opacity indicating system size.  In all three cases we observe the thermal density matrix has constant QFI density while the peak dynamical QFI scales at the Heisenberg limit. This illustrates the fundamental difference between between the metrological utility of equilibrium states vs quench dynamics. In the case of the Ising model, the dynamical QFI of the $Z_{\rm tot}$ operator is compared with the thermal QFI of the $X_{\rm tot}$ operator, since the thermal QFI of $Z_{\rm tot}$ is negligible.}
    \label{fig:therm-v-dyn}
\end{figure}

\begin{figure}
	\centering
 \includegraphics[width=0.99\textwidth]{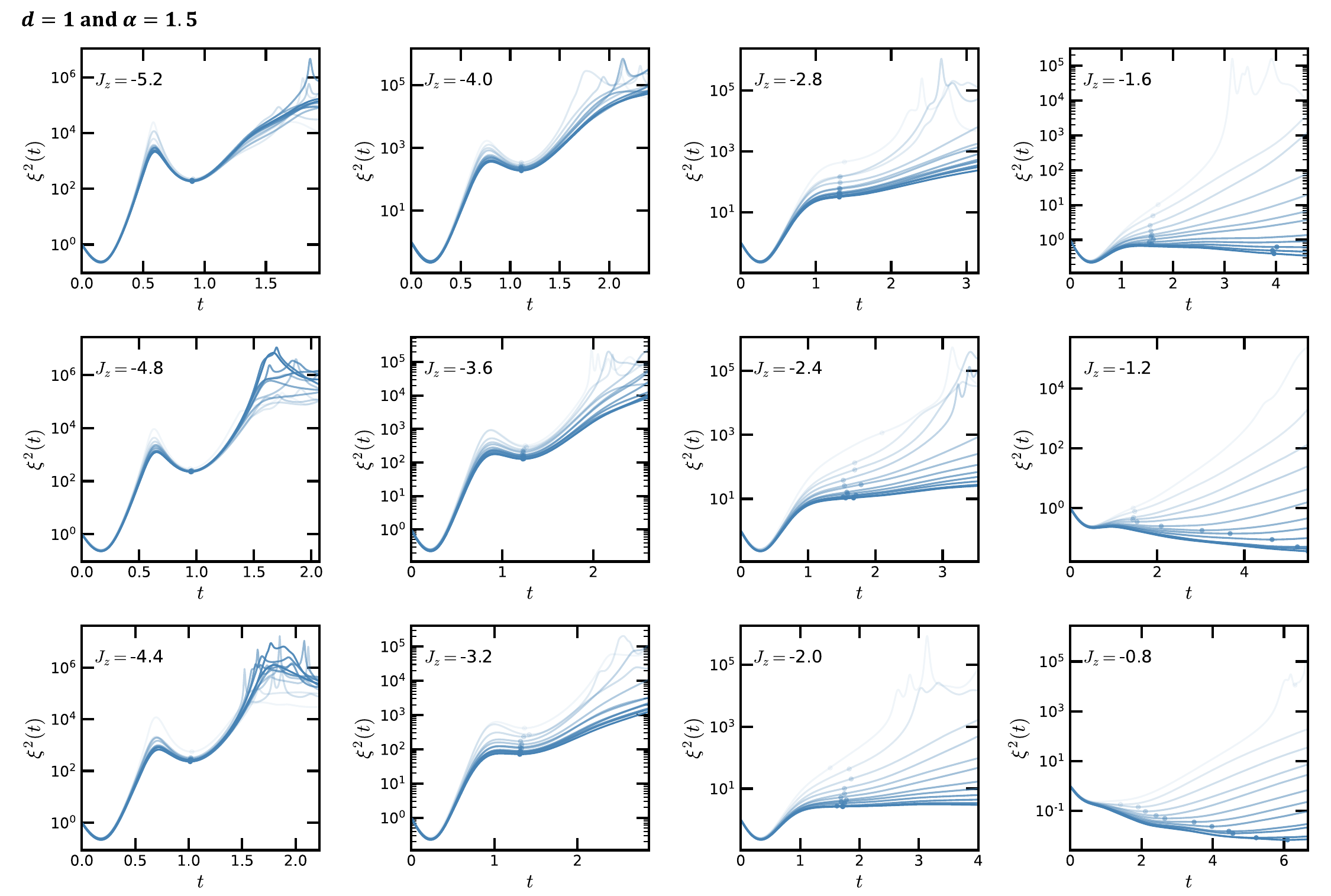}
	\caption{Squeezing as a function of time from DTWA simulations for $d=1$ $N=[120...10379]$. Opacity increases with system size. Circular markers indicate the characteristic squeezing $\xi^2_{\rm opt}$ determined via the technique described in the Methods.}
	\label{fig:1-1d}
\end{figure}

\begin{figure}
	\centering
 \includegraphics[width=0.99\textwidth]{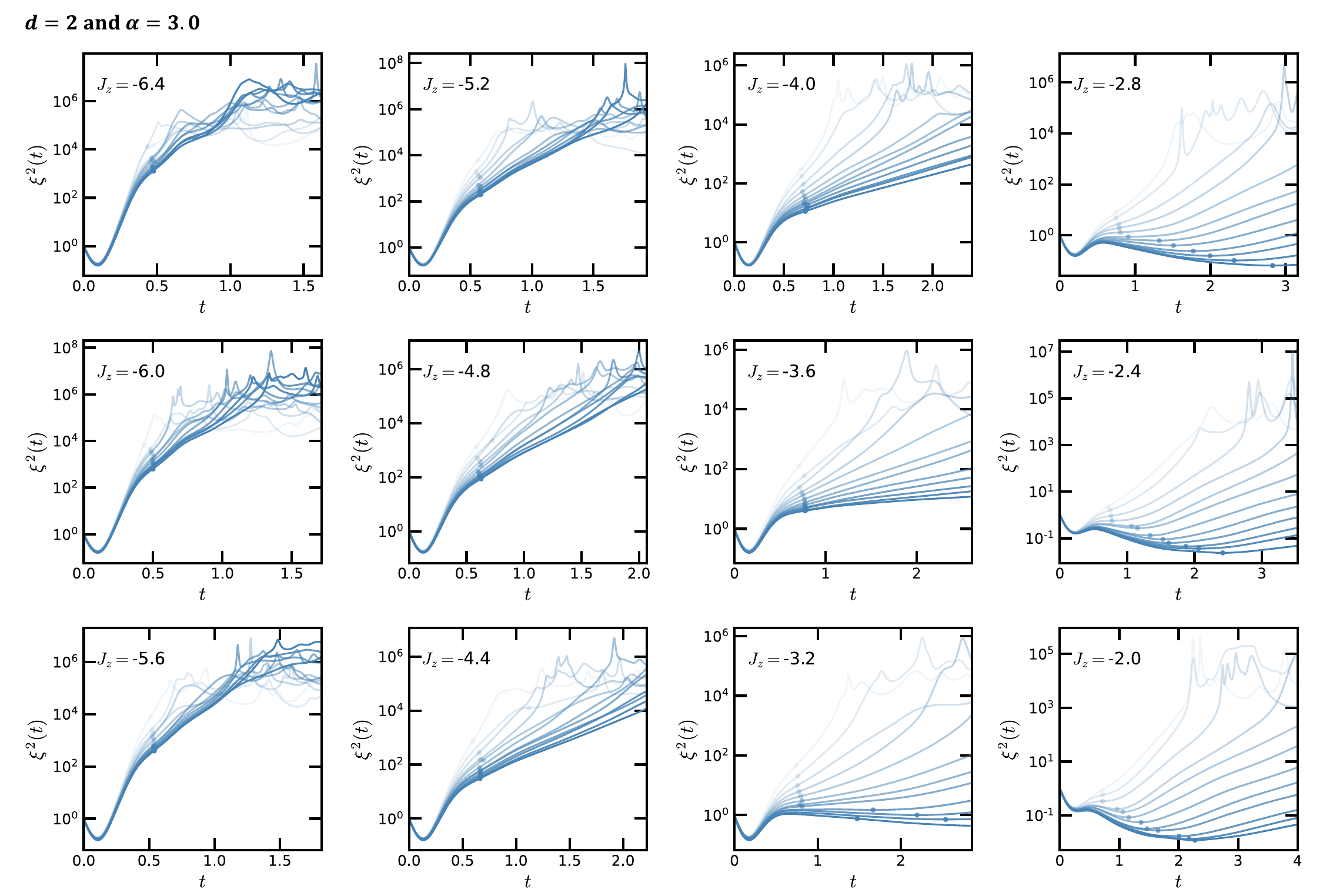}
	\caption{Squeezing as a function of time from DTWA simulations for $d=2$ $L=[11...102]$. Opacity increases with system size. Circular markers indicate the characteristic squeezing $\xi^2_{\rm opt}$ determined via the technique described in the Methods.}
	\label{fig:1-2d}
\end{figure}

\begin{figure}
	\centering
 \includegraphics[width=0.99\textwidth]{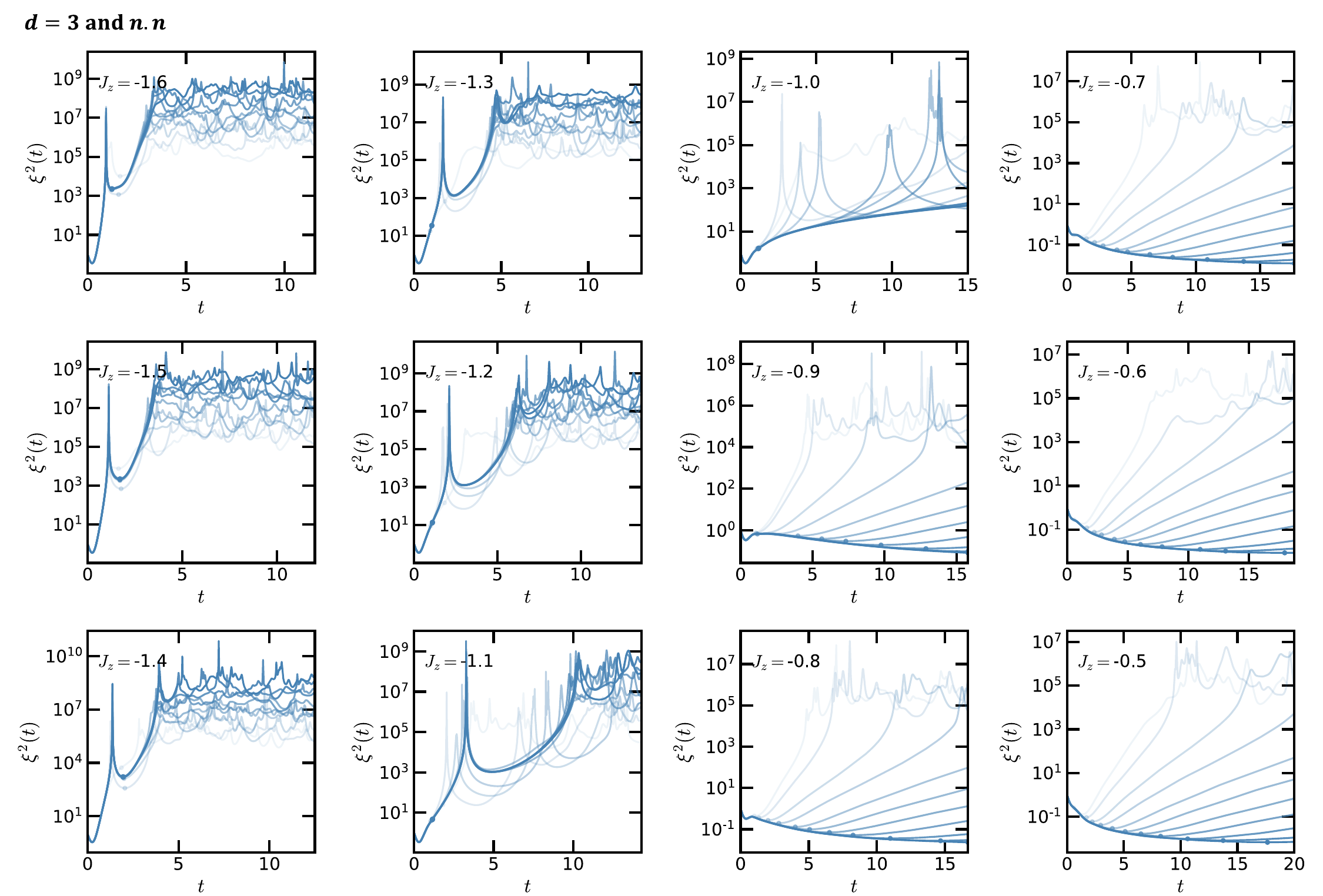}
	\caption{Squeezing as a function of time from DTWA simulations for $d=3$ with nearest-neighbor interactions for $L=[5...50]$. Opacity increases with system size. Circular markers indicate the characteristic squeezing $\xi^2_{\rm opt}$ determined via the technique described in the Methods.}
	\label{fig:1-3d}
\end{figure}



\begin{figure}
	\centering
 \includegraphics[width=0.99\textwidth]{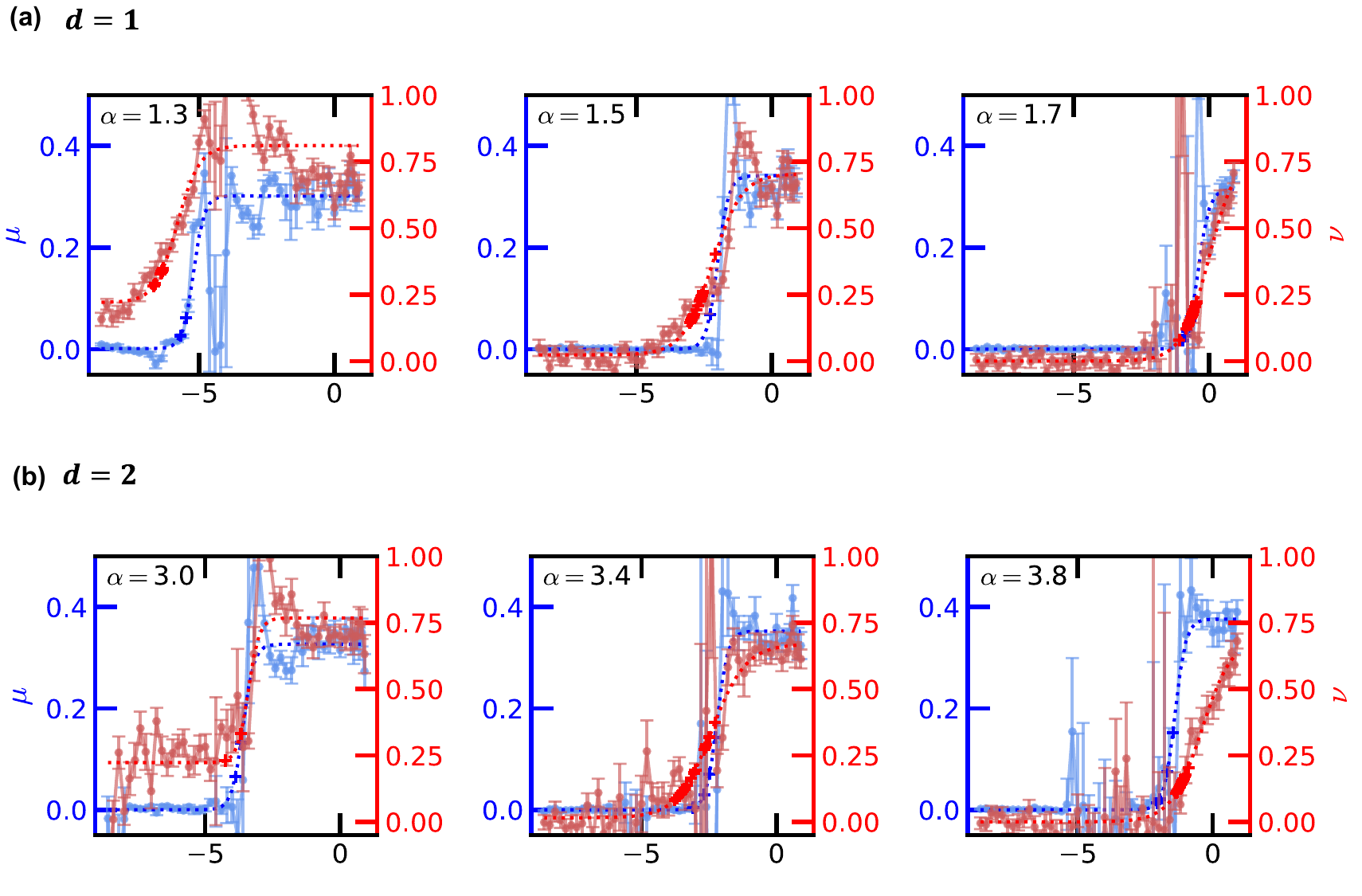}
	\caption{(a) Squeezing scaling exponents $\nu$ (red) and $\mu$ (blue) as functions of $J_z$ for varying $\alpha$ in $d=1$ presented as average values over 25 resamples of DTWA results $\pm$ SD. Estimates of the onset of each exponent are marked by ``$+$". The estimates are obtained by fitting $\mu(J_z)$, $\nu(J_z)$ for each resample with logistic curves. The left elbow of each logistic provides an estimate of the critical point. These estimates are superimposed on a logistic fit of average $\mu$, $\nu$. The estimates of the squeezing critical point fall in a broad but consistent region that tracks the thermal critical point (Fig.~1). (b) Analogous results for $d=2$.}
	\label{fig:3-1d}
\end{figure}

